\documentclass[useAMS,usenatbib]{mn2e}
\usepackage{graphicx}
\title{Curvature of the spectral energy distribution, the dominant process for inverse Compton component and other jet properties in Fermi 2LAC blazars}
\author[R. Xue et al.]{\\R. Xue$^{1}$, D. Luo$^{1}$, L. M. Du$^{1}$\thanks{E-mail: leiming\_du@ynao.ac.cn}, Z. R. Wang$^{1}$, Z. H. Xie$^{1}$, T. F. Yi$^{1}$, D. R. Xiong$^{2}$, Y. B. Xu$^{1}$, W. G. Liu$^{1}$ and X. L. Yu$^{1}$ \\
$^{1}$Department of Physics, Yunnan Normal University, Kunming 650500, China\\
$^{2}$National Astronomical Observatories/Yunnan Observatories, Chinese Academy of Sciences, Kunming 650011, China}
\begin{document}

\pagerange{\pageref{firstpage}--\pageref{lastpage}} \pubyear{2002}

\maketitle

\label{firstpage}

\begin{abstract}
We fit the spectral energy distributions (SEDs) of members of a large sample of Fermi 2LAC blazars to synchrotron and inverse Compton (IC) models. Our main results are as follows. (i) As suggested by previous works, the correlation between peak frequency and curvature can be explained by statistical or stochastic particle acceleration mechanisms. For BL Lacs, we find a linear correlation between synchrotron peak frequency and its curvature. The slope of the correlation is consistent with the stochastic acceleration mechanisms and confirm previous studies. For FSRQs, we also find a linear correlation, but its slope cannot be explained by previous theoretical models. (ii) We find a significant correlation between IC luminosity and synchrotron luminosity. The slope of the correlation of FSRQs is consistent with the EC process. And the slope of the correlation of BL Lac is consistent with the SSC process. (iii) We find several significant correlations between IC curvature and several basic parameters of blazars(black hole mass, broad line luminosity, the Lorentz factor of jet). We also find significant correlations between bolometric luminosity and these basic parameters of blazars which suggest that the origin of jet is a mixture of the mechanisms proposed by Blandford $\&$ Znajek and by Blandford $\&$ Payne.
\end{abstract}

\begin{keywords}
radiation mechanisms: nonthermal -- galaxies: active -- galaxies: jet -- BL Lacertae objects: general
\end{keywords}

\section{Introduction}

Blazars are the most extreme form of active galactic nuclei (AGN), with their jets pointed in the direction of the observer(Urry $\&$ Padovani 1995). The typical spectral energy distribution (SED) of blazars is generally described by a double bump structure. Harvey, Wilking $\&$ Joy (1982) were the first to propose using a smooth spectrum instead of a segmented power-law spectrum when they researched the submillimeter band of 3C 345. Landau (1983) researched a sample in the centimeter, millimeter and optical bands that consisted of 9 quasi-stellar objects(QSOs) and BL Lacs. Landau et al. (1986) analyzed a sample of low-energy peaked BL Lac objects from the millimeter band to the ultraviolet band, and found that, although they could obtain spectrum successfully when combined with power-law spectrum, the data from the radio to the ultraviolet band showed good consistency with the smooth spectrum many cases. They used a quadratic function, namely:
\begin{equation}
log S_{\nu}=C+[(log \nu -B)^2]/2A
\end{equation}
to fit the spectrum. Massaro et al. (2004a) found that the log-parabolic law can well describe the X-ray peak value of SED of Mrk 421, and subsequent found that log-parabolic law can fit the X-ray spectrum(Massaro et al. 2008).

There is, however, a problem in using the log-parabolic law to fit SEDs: some parameters in the equation do not have a physical explanation. It is remarkable that the SEDs of many sources can be fitted by a quadratic function that contains three parameters. The property indicates that these sources may have a similar structure, or similar relativistic particle energy distributions, or both(Landau et al. 1986). With a more general and simple situation about the particles increasing their energy, the log-parabolic law is used to statistical acceleration structure. One advantage of log-parabolic law is that it can be used to calculate various useful parameters, such as peak frequency, curvature and so on, which is simpler than the case for other models.

When we obtain the value of peak frequency and peak luminosity of SEDs, the curvature of the SEDs will be another important parameter. The curvature is the property possessed by the curving of a line. If the peak frequency and peak luminosity are known, the curvature can be used to derive the bolometric luminosity. The relationship between the peak frequency and the curvature can be explained in terms of the acceleration process of emitting electrons(Massaro et al. 2004a, 2004b, 2006; Paggi et al. 2009a, 2009b; Rani et al. 2011; Tramacere et al. 2007, 2009, 2011; Chen 2014). Chen (2014) performed an analysis of the curvature-peak frequency connection, and re-derived a theoretical model that some authors had used before(Massaro et al. 2004a, 2006; Tramacere et al. 2011). He predicted two electron acceleration mechanisms according to the coefficient of the curvature-peak frequency relationship. For models of stochastic acceleration, the values of the energy-dependent acceleration probablility, the fluctuation of the fractional acceleration gain(the latter two are statistical acceleration) and the slope k(1/-c=k$log{\nu}_{p}+h$) are 2, 2.5, 3.333 respectively. The result of Chen (2014) was 2 which is consistent with stochastic acceleration.

The SEDs generated by two emission components, namely the synchrotron component and inverse Compton(IC) component (Ghisellini et al. 1997; Massaro et al. 2004a, 2006). In the lepton model, it is generally accepted that low-energy peak is caused by the synchrotron emission of relativistic electrons in the jet, and that the high energy peak is caused by IC scattering (Massaro et al. 2004, 2006; Meyer et al. 2012). There is, however, disagreement concerning the origin of the soft photons of IC scattering. (1) They are derived from synchrotron emission, termed the synchrotron self-Compton(SSC) process(Rees 1967; Jones et al. 1974; Marscher $\&$ Gear 1985; Maraschi et al. 1992; Sikora et al. 1994; Bloom $\&$ Marscher 1996). (2) They are derived from the exterior of jets, termed the external Compton(EC) process. There are three possible sources of EC soft photons: accretion disk photons entering jets directly(Dermer et al. 1992; Dermer $\&$ Schlickeiser 1993); broad line region (BLR) photons entering jets(Sikora et al. 1994; Dermer et al. 1997); and dust torus infrared radiation photons entering jets (Blazejowski et al. 2000; Arbeiter et al. 2002). Ghisellini (1996) derived two relationships between the synchrotron luminosity and the IC luminosity so that it could be determined whether the IC component is dominated by the EC process or the SSC process($L_{EC}$$\sim$$L_{syn}^{1.5}$, $L_{SSC}$$\sim$$L_{syn}^{1.0}$).

The bolometric luminosity is one of the most important parameters of blazars. It represents the amount of electromagnetic energy a body radiates per unit of time. Thus it represents the total radiant energy over a wide band (from the radio band to the $\gamma$ band). In research concerning the origin of jets, the current theoretical model is that the jet power is generated from accretion and the extraction of rotational energy or angular momentum from disc/black hole(Blandford $\&$ Znajek 1977; Blandford $\&$ Payne 1982). The black hole and accretion disc play an important role in the process, so it is important to research the relationships between the jet and black hole and between the jet and accretion. Ghisellini et al. (2010) found that Fermi blazars with higher luminosities may have a larger black hole. Xiong et al. (2014a) researched the relationship between the jet power and black hole mass and found that there is a significant correlation between them, which means that the the jet power is controlled by the spin of black hole. In both mechanisms, the magnetic field plays a major role in channelling power from the black hole or the disc into the jet. The process be sustained by matter accreting on to black hole, so it is reasonable that there is connection between the origin of jet and accretion(Maraschi $\&$ Tavecchio 2003). Many authors have studied the jet-accretion disk relationship, using a variety of methods(Rawlings $\&$ Saunders 1991; Falcke $\&$ Biermann 1995; Serjeant et al. 1998; Cao $\&$ Jiang 1999; Wang et al. 2004; Liu et al. 2006; Xie et al. 2007; Gu et al. 2009; Ghisellini et al. 2009a, 2009b, 2010, 2011; Sbarrato et al. 2012; Yu et al. 2015). The BLR luminosity can be taken as an indicator of the accretion power of blazars (Celotti, Padovani $\&$ Ghisellini 1997), and the bolometric luminosity can be taken as the index of jet power, so researching the relationship between the BLR luminosity and bolometric luminosity can provide information on the relationship between the jet and accretion disc, contributing to the knowledge about the origin of jets. Xie et al. (2007) found a significant correlation between $log L_{BLR}$ and $log L_{jet}$($log L_{BLR}$=$log L_{jet}$+$log {\eta}$+const). The Lorentz factor ($\Gamma$) can describe the speed of the jet flow (Hovatta et al. 2009). There is a relationship between the jet power and jet speed: more powerful jets will appear to be faster (Kharb et al. 2010). L$\ddot{u}$ et al. (2012) found that the $\Gamma$ and $\gamma$-luminosity are significant connection. We know that the $\gamma$-luminosity can represent the bolometric luminosity(Fan, Xie, $\&$ Bacon 1999, Xie et al. 2004), so we can predict that there will be a linear correlation between $\Gamma$ and bolometric luminosity.

In this paper, the SEDs of both synchrotron and IC components of a sample are fitted by a log-parabolic law. The bolometric luminosity is the amount of electromagnetic energy a body radiates per unit of time,it is the total radiant energy over a wide band (from the radio band to the $\gamma$ band), we calculate the exact value of the bolometric luminosity and curvature by fitting the SEDs. Firstly, we analyse the curvature-peak frequency relation and try to verify the particle acceleration mechanism. Then, we analyse the correlation between the IC luminosity and the synchrotron luminosity and try to judge the IC component is dominated by the EC process or the SSC process. Finally, we analyse the correlations between curvature, bolometric luminosity and black hole mass, BLR luminosity and the Lorentz factor.

The paper is structured as follows. In Section 2 we present the sample; In Section 3 we detail the fitting procedure; in Section 4, we present the results; In Section 5 we provide a discussion and conclusion. A $\Lambda$CDM cosmology with $H_{\rm{0}}={\rm{70Km s^{-1} Mpc^{-1}}}$, $\Omega_{\rm{m}}=0.27$, $\Omega_{\rm{\Lambda}}=0.73$ is adopted.

\section{The sample}

We collected a sample from the second LAT AGN catalog(2LAC). At least one of the two components(synchrotron and IC component) of the blazars in our sample can be fitted by sufficient multifrequency data coverage. The sample contains 279 blazars, including 200 flat-spectrum radio quasars(FSRQs) and 79 BL Lacs. For the FSRQs, the IC components of 98 objects and the synchrotron components of 21 FSRQs cannot be fitted, while the complete SEDs of 81 FSRQs can be fitted. For the BL Lacs, the IC components of 46 objects and the synchrotron components of 5 objects cannot be fitted, while the complete SEDs of 28 BL Lacs can be fitted. According to Abdo et al. 2010, blazars can be subdivided into three subclasses, namely LSP(low synchrotron-peaked, ${\nu}_{peak}^{syn}$$<10^{14}$Hz), ISP(intermediate synchrotron-peaked, $10^{14}Hz<$${\nu}_{peak}^{syn}$$<10^{15}$Hz) and HSP(high synchrotron-peaked, ${\nu}_{peak}^{syn}$$>10^{15}$Hz). For FSRQs, 179 (89.5 percent of the total) have an SED classification, namely 175 LSPs and 4 ISPs. For BL Lacs, 74(93.7 percent of the total) have an SED classification, namely 37 LSPs, 11 ISPs and 26 HSPs.

For our sample, biases are as follows:

(1) For blazars in 2LAC, 526 had an SED classification, with LSP representing the lagerst subclass(282/526=54 percent)(Ackermann et al. 2011). Our sample is a subsample of 2LAC, and 83.8 percent(212/253=83.8 pecent) are LSPs. Compared with 2LAC, our sample is more dominated by LSPs.

(2) Fermi data were integrated over a few months(Abdo. et al. 2010, Giommi et al. 2012a), and therefore the SED fitting, luminosity calculation and correlation analysis concerning the $\gamma$-ray band in this paper cannot be considered simultaneous in this paper. However, at least the fitting of synchrotron component can be considered simultaneous. The observation date of the (quasi-) simultaneous data of blazars whose IC components can be fitted is in the period of the Fermi exposure time.

(3) Redshift can be measured with a clear spectrum. However, BL Lacs typically lack some emission lines, so the redshift will not be accurate. This lack can affect the rest-frame correction, the luminosity calculation and the related correlation analysis(a detailed discussion is given in Sect. 5.1).

Detailed information about the sample is given in Table 5, with the following headings: column (1), name of Fermi catalog; column (2), redshift; column (3), the observation date of the (quasi-)simultaneous data; column (4), logarithm of the synchrotron peak frequency in units of Hz; column (5), second degree term of log-parabolic for the synchrotron component and the IC component; column (6),  logarithm of the black hole mass in the unit of $M_{\odot}$; column (7), logarithm of the broad-line luminosity in the units of erg $s^{-1}$; column (8), Lorentz factor of jet; column (9),  logarithm of luminosity for the synchrotron component and IC component in the units of erg $s^{-1}$; column (10), logarithm of the bolometric luminosity in the units of erg $s^{-1}$; column (11), type of the blazar.

For BL Lacs without a measured redshift, we assume the mean redshift value of 0.27 in 2LAC(Ackermann et al. 2011). For the 95 blazars for which more than one black hole masses is given in the literature, we use the average black hole mass instead in the related correlation analysis. All the values of luminosity are integrated from the SEDs.

\section{The fitting procedure}

We construct the SEDs of all the blazars in our samples from the multifrequency data using the ASDC SED Builder, an online service developed at the ASI Science Data Center(Stratta et al. 2011).
We use the second-degree polynomial function£º
\begin{equation}
log(\nu F_{\nu})=c(log \nu)^2+b(log \nu)+a
\end{equation}
to fit the synchrotron component and IC component separately, so that we could calculate the curvature around the peak(where the curvature is represented by $|2c|$). In this paper, all values are converted into the rest-frame. The parameters in the rest-frame of the second-degree polynomial function and the peak frequency can be calculated as:
\begin{equation}
a_{rest}=a-b\Delta+c\Delta^{2}+log4{\pi}+2logD_{L}
\end{equation}
\begin{equation}
b_{rest}=b-2c\Delta
\end{equation}
\begin{equation}
c_{rest}=c
\end{equation}
\begin{equation}
{\nu}_{p, rest}={\nu}_{p}^{obs}(1+z)
\end{equation}
, where $D_{L}$ is the luminosity distance, z is the source redshift and $\Delta$=$log(1+z)$.
The database of the ASDC SED Builder provides the observational data from many space telescopes and ground-based telescopes, and we can obtain simultaneous data based on the observation times provided. The IC components of some blazars(e.g. BL Lac-HSPs) covered only the $\gamma$-ray band, and therefore the IC-fitted parameters may be unreliable because of the narrow data coverage. In these cases we did not provide fits. According to the synchrotron and IC fitting curves, we can calculate the bolometric flux as follows:
\begin{equation}
F_{integrated}=\int F_{\nu}d{\nu}=(ln10)\int 10^{(cx^2+bx+a)} dx
\end{equation}
The bolometric luminosity can be calculated using the relationship:
\begin{equation}
L_{bol}=4\pi D_{L}^2 F_{integrated}.
\end{equation}

The biases and uncertainties that might be caused by the above procedure are as follows:

1. The thermal radiation from the disc/torus and the host galaxy is prominent in both the optical/IR and the UV wavebands(Ackermann et al. 2015, Giommi et al. 2012a, Abdo et al. 2010, Marscher 2009). Before fitting the SED, we had to exclude the thermal radiation because it will amplify the emission from the jets of blazars. We identified the thermal component by visual inspection, so we can separate the radiation whose flux is significantly greater than the non-thermal radiation at ultraviolet frequencies.

2. In this paper, the SEDs are fitted using a second-degree polynomial. However, the second-degree polynomial cannot handle asymmetry very well and may skew some relevant results(for example in the luminosity calculation). In order to evaluate this effect, we used four typical BL Lacs with an apparent asymetrical synchrotron component, namely 2FGLJ0050.6-0929, 2FGLJ0136.5+3905, 2FGLJ1015.1+4925 and 2FGLJ1136.7+7009, and fitted their synchrotron component with a second-degree polynomial and with a third-degree polynomial and then compare the synchrotron luminosities. For the second-degree polynomial, the logarithms of synchrotron luminosity were 47.2, 46.8, 46.1 and 44.6. For the third-degree polynomial, the synchrotron luminosity were 46.6, 46.1, 45.2 and 44.0. From the above results, we found that the logarithm of synchrotron luminosity integrated by the third-degree polynomial is lower than that from the second-degree polynominal, and the numerical differences are within 1. We therefore suggest that the impact on calculation of luminosity is insignificant, and will not affect the main results in this paper.

3. The light variability of blazars can cause massive changes in the SED curve, peak frequency, peak flux, curvature, etc(Massaro et al. 2004a, Massaro et al. 2004b, Acciari et al. 2011). In this paper, we make sure that the fitted data were observed contemporaneously. The influence of light variability is not considered here. In order to
make use of the maximum availability of (quasi-)simultaneous data coverage, SEDs in both low and high states are considered. This might contribute to the scatter in the correlation analysis below

\section{The results}

Three sub-samples selected from our sample were analysed using linear correlations. The sub-samples are as follows:

(A) FSRQs(N=179, where N is number of objects in the sample) and BL Lacs(N=74) with fitted synchrotron components. We studied the correlation between synchrotron peak frequency and its curvature using this subsample.

(B) FSRQs(N=81) and BL Lacs(N=28) with fitted complete SEDs. We studied the correlation between the IC luminosity and synchrotron luminosity using this subsample.

(C) FSRQs(N=59) with black hole masses, broad line luminosity, Lorentz factor and fitted complete SEDs. This is a clean sample for FSRQs.
For the three subsamples, biases are as follows:

(1) We used sub-samples A and B for the following reasons. First, they enable the study of interesting correlations that are important for FSRQs and BL Lacs. Second, the statistical significance is maximized because they provide the maximum availability of data. As noted above, these two subsamples predominantly contain objects with low peaks. This means that they might give biased results. The subsamples could give more reliable results if they contained more high-peaked BL Lacs. Fortunately, although the redshifts of BL Lacs are not measured accurately, these two correlations are not affected seriously.

(2) Only 59 FSRQs(59/281=21 percent) get a complete SED and data for the black hole mass, broad-line luminosity and Lorentz factor. The rest of the sample do not have all the data(black hole mass, broad-line luminosity or the Lorentz factor) or a complete SEDs. The missing data may have an enormous impact on the correlation analysis, and it is difficult to assess what has happened to the sample between each plots. Therefore, the results that we obtain are only for our subsample. A larger clean sample of FSRQs is needed to check these results in detail.

\subsection{Synchrotron peak frequency vs synchrotron curvature in FSRQs and BL Lacs}

The synchrotron peak frequency versus synchrotron curvature is plotted in Fig. 1 using data from subsample A.. Here we use the -1/$c_{syn}$ instead of 2($-c_{syn}$) to represent the synchrotron curvature because it will be convenient to compare with the theoretical results(see Chen 2014). The Spearman test was used to analyse the correlation between $log {\nu}_{peak}^{syn}$ and -1/$c_{syn}$ for FSRQs(top panel) and BL Lacs(bottom panel). We assume that the correlation is significant when $p<0.05$. For FSRQs, the Spearman test gives a significance level $p<0.0001$ and a coefficient of correlation R=0.641. The bisector linear regression gives the best linear fitting equation as -1/$c_{syn}$=(3.69$\pm$0.24)$log {\nu}_{peak}^{syn}$+(-42.57$\pm$3.14). For BL Lacs, the Spearman test gives a significance level $p<0.0001$ and a coefficient of correlation R=0.702. The bisector linear regression gives the best linear fitting equation as -1/$c_{syn}$=(1.87$\pm$0.19)$log {\nu}_{peak}^{syn}$+(-19.21$\pm$2.76). The results show that the correlations between $log {\nu}_{peak}^{syn}$ and -1/$c_{syn}$ for FSRQs and BL Lacs are significant. For BL Lacs, the slope of the correlation($k_{BL Lac}$=1.87$\pm$0.19) is consistent with the stochastical acceleration mechanism(Chen 2014). For FSRQs, however, the slope($k_{FSRQ}$=3.69$\pm$0.24) is not consistent with any of the theoretical values provided by Chen (2014). The distribution(the bottom panel of Fig. 1) of BL Lac-ISPs and BL Lac-HSPs is more dispersed than the distribution of BL Lacs-LSPs. If the number of BL Lac-ISPs and BL Lac-HSPs were greatly increased, this may change the slope of the best linear fit and the significant level.

\begin{figure}
\includegraphics[width=84mm]{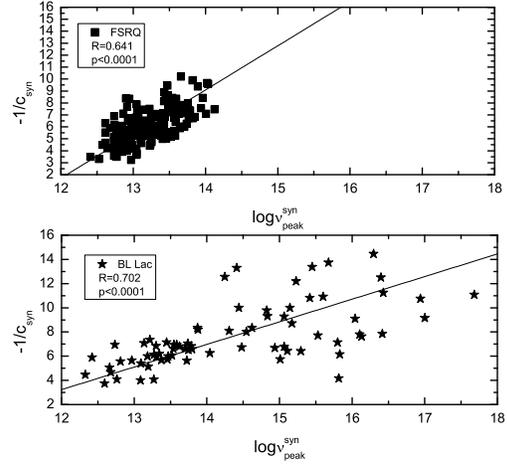}
\caption{Top panel: The correlation between $log {\nu}_{peak}^{syn}$ and -1/$c_{syn}$ for FSRQs. The solid line is the best linear fitting($p<0.0001$).
Bottom panel: The correlation between $log {\nu}_{peak}^{syn}$ and -1/$c_{syn}$ for BL Lacs. The solid line is the best linear fitting($p<0.0001$).}
\end{figure}

\subsection{Inverse Compton luminosity vs synchrotron luminosity in FSRQs and BL Lacs}

The inverse Compton luminosity versus synchrotron luminosity is plotted in Fig. 2 using data from subsample B. The Spearman test was applied in order to analyse the correlation between $log L_{IC}$ and $log L_{syn}$ for FSRQs(top panel) and BL Lacs(bottom panel). For FSRQs, the Spearman test gives a significance level $p<0.0001$ and a coefficient of correlation R=0.74. The bisector linear regression gives the best linear fitting equation as $log L_{IC}$=(1.45$\pm$0.11)$log L_{syn}$+(-20.10$\pm$5.12). For BL Lacs, the Spearman test gives a significance level $p<0.0001$ and a coefficient of correlation R=0.901. The bisector linear regression gives the best linear fitting equation as $log L_{IC}$=(1.12$\pm$0.10)$log L_{syn}$+(-5.43$\pm$4.61). The results show that the correlations between $log L_{IC}$ and $log L_{syn}$ for FSRQs and BL Lacs are significant, and $k_{FSRQ}$=1.45$\pm$0.11, $k_{BL Lac}$=1.12$\pm$0.10. According to Ghisellini (1996)($L_{EC}$$\sim$$L_{syn}^{1.5}$,$L_{SSC}$$\sim$$L_{syn}^{1.0}$), our results suggest that the IC component of FSRQs is dominanted by the EC process(Ghisellini et al. 2002; Celotti $\&$ Ghisellini 2008; Ghisellini et al. 2010; Finke 2013) and that in BL Lacs it is dominanted by the SSC process(Zhang et al. 2013; Lister et al. 2011; Celotti $\&$ Ghisellini 2008; Ghisellini et al. 2002; Ghisellini et al. 2010; Ackermann et al. 2012). In subsample B, the number of BL Lacs is much lower than the number of FSRQs, and a larger sample of BL Lacs with complete SEDs is needed for further study.

\begin{figure}
\includegraphics[width=84mm]{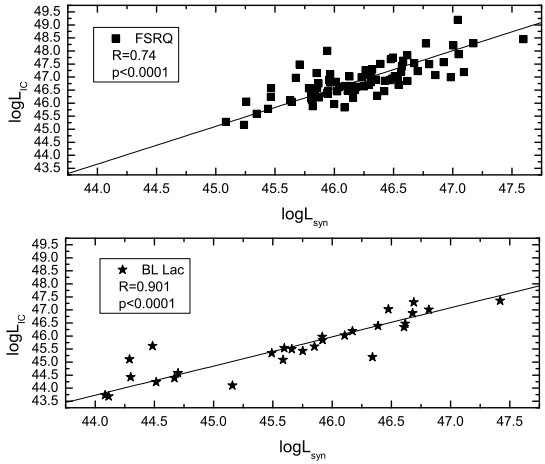}
\caption{Top panel: The correlation between $log L_{IC}$ and $log L_{syn}$ for FSRQs. The solid line is the best linear fitting($p<0.0001$).
Bottom panel: The correlation between $log L_{IC}$ and $log L_{syn}$ for BL Lacs. The solid line is the best linear fitting($p<0.0001$).}
\end{figure}

\subsection{Broad line luminosity, black hole mass, the Lorentz factor and bolometric luminosity vs curvature in FSRQs}

The broad-line luminosity, black hole mass, Lorentz factor and bolometric luminosity are plotted versus IC curvature in Figs 3 to 6 using data from subsample C. It can be seen in these figures that 2FGL J1539.5+2747 and 2FGL J2211.9+2355 could be outliers because of their excessive IC curvature. We checked their fitted SEDs by eye so that fitting errors can be excluded. We also found that the Lorentz factor of 2FGL J1733.1-1307($\Gamma$=65.24) is extremmely high. According to Hovatta et al. (2009), either the Lorentz factor is accurate, or the source exhibits so rapid flares that the fast variations were undetected in the monitoring programmes. For this reason, we exclude 2FGL J1733.1-1307 in the following analysis concerning the Lorentz factor, as did Hovatta et al. (2009).

The Spearman test was applied in order to analyse the correlations between the broad-line luminosity, black hole mass, Lorentz factor, bolometric luminosity and IC curvature for FSRQs(there are no correlations between synchrotron curvature and all the parameters). The results are as follows:

1. $log L_{BLR}$ vs. -1/$c_{IC}$: $p<0.0001$, R=0.496.

2. $log M_{BH}$ vs. -1/$c_{IC}$: $p=0.003$, R=0.384.

3. $\Gamma$ vs. -1/$c_{IC}$: $p=0.001$, R=0.418.

4. $log L_{bol}$ vs. -1/$c_{IC}$: $p=0.001$, R=0.417.

The results show that the correlations between all the parameters and the IC curvature for FSRQs are moderately significant. The significant correlation between $log L_{BLR}$ and -1/$c_{IC}$ can be explained as the origin of the soft photons of the IC component may be dominated by the broad-line region.

\begin{figure}
\includegraphics[width=84mm]{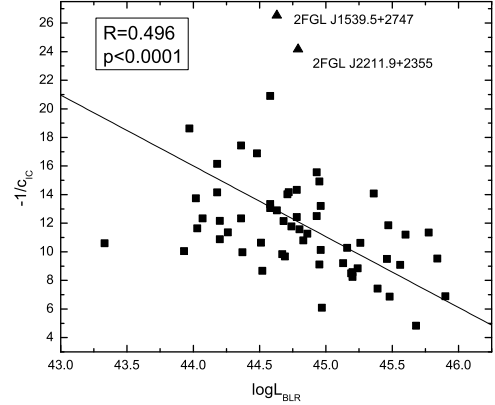}
\caption{The correlation between $log L_{BLR}$ and -1/$c_{IC}$ for FSRQs. The solid line is the best linear fitting($p<0.0001$).}
\end{figure}

\begin{figure}
\includegraphics[width=84mm]{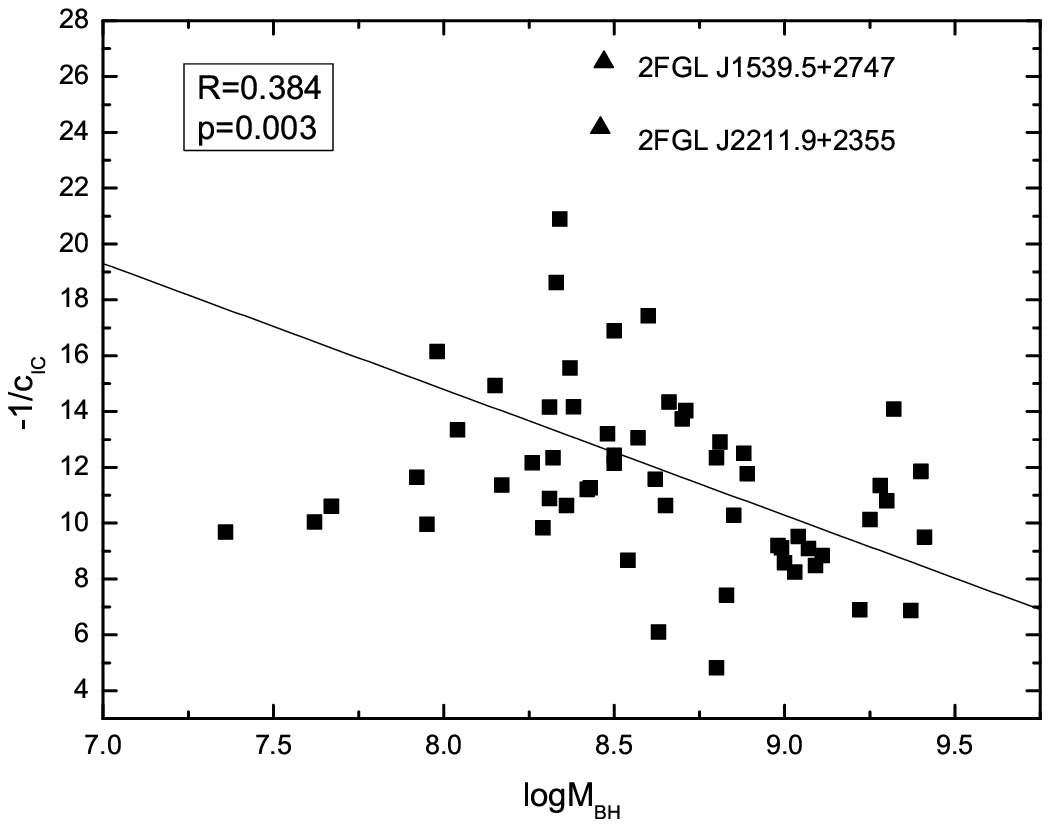}
\caption{The correlation between $log M_{BH}$ and -1/$c_{IC}$ for FSRQs. The solid line is the best linear fitting($p=0.003$).}
\end{figure}

\begin{figure}
\includegraphics[width=84mm]{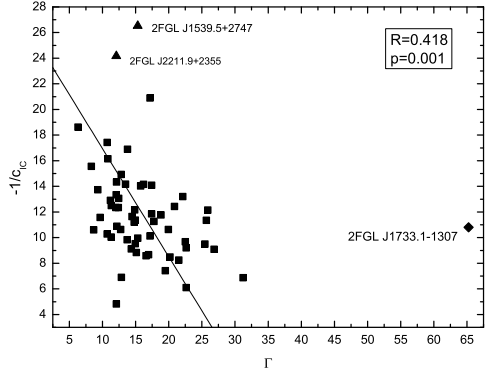}
\caption{The correlation between $\Gamma$ and -1/$c_{IC}$ for FSRQs. The solid line is the best linear fitting($p=0.001$).}
\end{figure}

\begin{figure}
\includegraphics[width=84mm]{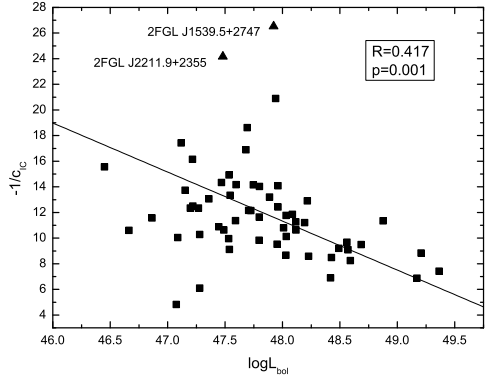}
\caption{The correlation between $log L_{bol}$ and -1/$c_{IC}$ for FSRQs. The solid line is the best linear fitting($p=0.001$).}
\end{figure}

\subsection{Broad line luminosity, black hole mass and the Lorentz factor vs bolometric luminosity in FSRQs}

The broad-line luminosity, black hole mass, and Lorentz factor are plotted versus bolometric luminosity in Figs 7 to 9 using data from subsample C. As noted above, the Lorentz factor of 2FGL J1733.1-1307 is excluded as an outlier. The Spearman test was applied in order to analyse the correlations between the broad line luminosity, black hole mass, Lorentz factor and bolometric luminosity for FSRQs. The results are as follows:

1. $log L_{BLR}$ vs. $log L_{bol}$: $p<0.0001$, R=0.575

2. $log M_{BH}$ vs. $log L_{bol}$: $p<0.0001$, R=0.47

3. $\Gamma$ vs. $log L_{bol}$: $p<0.0001$, R=0.767

The results show that the correlations between all the parameters and $log L_{bol}$ for FSRQs are significant. The fact that the correlations between $log L_{BLR}$, $log M_{BH}$ and $log L_{bol}$ are significant support the theory that the origin of jet is a mixture of the mechanisms proposed by Blandford $\&$ Znajek and by Blandford $\&$ Payne(Punsly $\&$ Coroniti 1990; Meier et al. 1999, 2001). Furthermore, the significant correlation between $\Gamma$ and $log L_{bol}$ means that more powerful jets, the plasma blob in jets will move faster(Kharb et al. 2010; Wu et al. 2011; L$\ddot{u}$ et al. 2012).

\begin{figure}
\includegraphics[width=84mm]{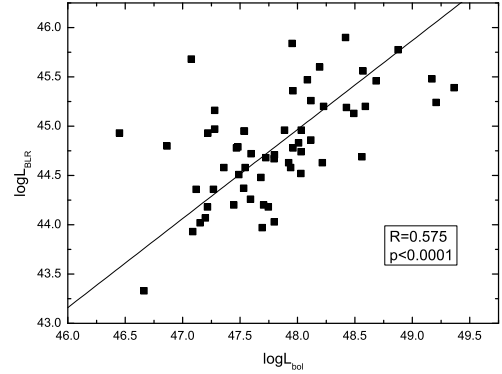}
\caption{The correlation between $log L_{BLR}$ and $log L_{bol}$ for FSRQs. The solid line is the best linear fitting($p<0.0001$).}
\end{figure}

\begin{figure}
\includegraphics[width=84mm]{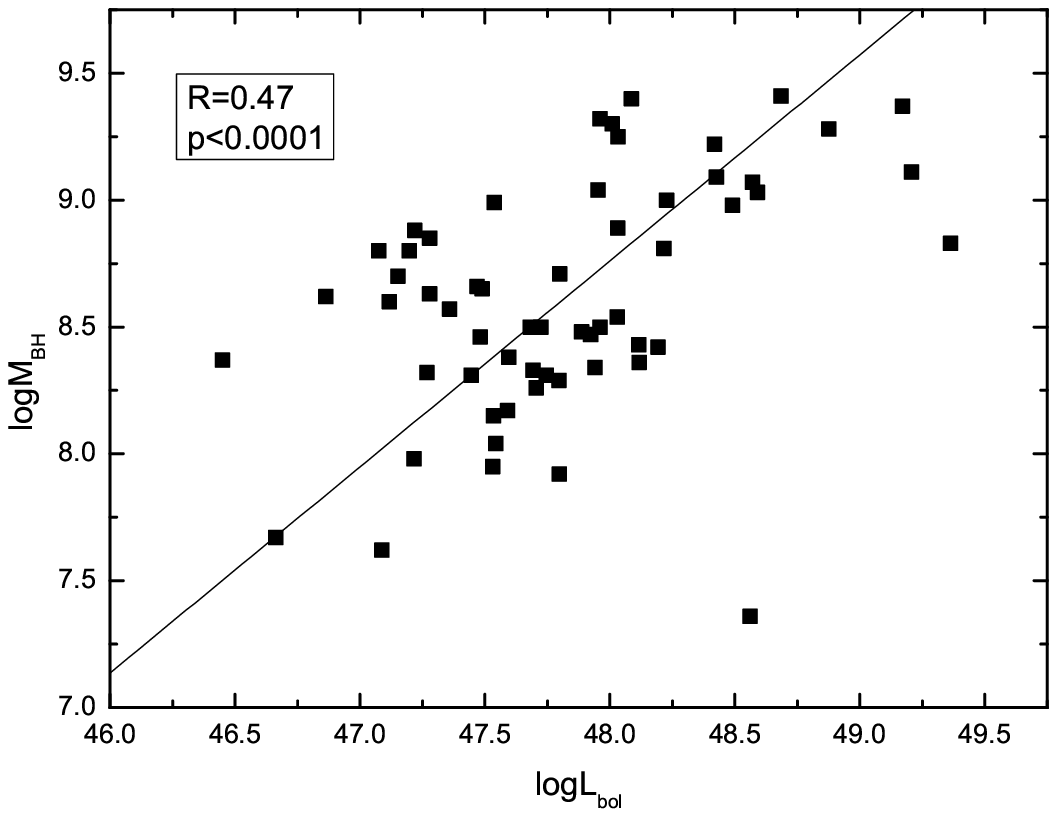}
\caption{The correlation between $log M_{BH}$ and $log L_{bol}$ for FSRQs. The solid line is the best linear fitting($p<0.0001$).}
\end{figure}

\begin{figure}
\includegraphics[width=84mm]{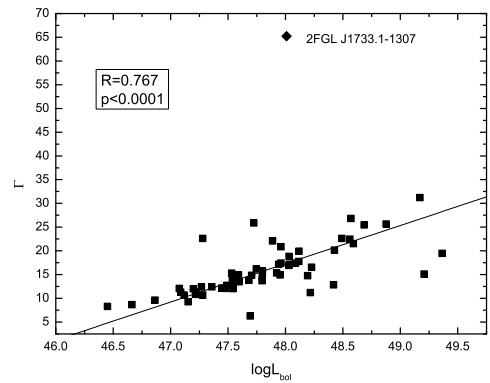}
\caption{The correlation between $\Gamma$ and $log L_{bol}$ for FSRQs. The solid line is the best linear fitting($p<0.0001$).}
\end{figure}

\section{Discussions and Conclusion}

\subsection{The effect of the inaccurate redshift of BL Lacs in related linear correlation analysis}

Redshifts are normally obtained by spectroscopic and photometric measurement. However, BL Lacs typically lack emission lines. Clear spectroscopic measurements can be obtained for only a small proportion of BL Lacs. Most redshifts of BL Lacs are obtained by photometric measurement or are based on dubious private communications. In Plotkin et al. (2008), photometric redshifts were estimated to be accurate to $\Delta$z$\approx$0.01 from comparison with spectroscopic redshifs. However, the two types of redshift measured for a few BL Lacs have a huge difference, $\Delta$z$\approx$0.5. The use of inaccurate redshifts may have a significant effect on the calculation of the luminosity and peak frequency in rest-frame.

In this paper, the inaccurate redshifts may affect the correlation between $log {\nu}_{peak}^{syn}$ and -1/$c_{syn}$ and between $log L_{IC}$ and $log L_{syn}$, discussed in Sect. 4.1 and Sect. 4.2, respectively. In subsamples A and B, there are 17 BL Lacs without measured redshifts. Therefore, we discuss the effect of the inaccurate redshift for two scenarios: (i) all BL Lacs in subsamples A and B. (ii) BL Lacs with measured redshifts in subsamples A and B. On the basis of the redshifts in our sample, we add a normally distributed disturbance. The mean value of this normally distributed disturbance(${\mu}_{disturbance}$) are 10\%z, 20\%z, \textbf{30\%z, 40\%z and 50\%z}, respectively. The standard deviation($\sigma$) are 10\%$\times$0.3144, 20\%$\times$0.3144, 30\%$\times$0.3144, 40\%$\times$0.3144 and 50\%$\times$0.3144(0.3144 is the standard deviation of the redshift distribution of BL Lacs in 2LAC). We can then obtain random redshifts and calculate $log L_{syn}$, $log L_{IC}$ and $log {\nu}_{peak}^{syn}$ in rest-frame. Finally, we repeated the linear correlations 5000 times and analysed the distributions of the coefficient of correlation R, the significance level p and slope K. The results are given in Table 1 to 4.

Table 1 and 2 show the distributions of R, p and K for the correlation between $log {\nu}_{peak}^{syn}$ and -1/$c_{syn}$ for all BL Lacs and for BL Lacs with a measured redshift in subsample A. From Tables 1 and 2, it can be seen that all p-values are less than 0.0001 and that all the distributions of R are normally distributed. The mean values of R(${\mu}_{R}$) in Table 1 are almost 0.701 and those of ${\mu}_{R}$ in Table 2 are almost 0.66. Furthermore, all the standard deviation of R(${\sigma}_{R}$) are small, ${\sigma}_{R}$$\leq$0.02. It can also be seen that all the distributions of K are normal. Their mean values(${\mu}_{K}$) are 1.87 and 1.89, respectively. Their standard deviations are all small, ${\sigma}_{K}$$\leq$0.01. Table 3 and 4 show the distributions of R, p and K for the correlation between $log L_{IC}$ and $log L_{syn}$ for all BL Lacs and for BL Lacs with measured redshift in subsample B. From these two table, it can be seen that almost all the p-values are less than 0.0001, but the distributions of R are not normal any more. In the correlation between $log L_{IC}$ and $log L_{syn}$, the coefficient of R is 0.901. Therefore, we want to know how many R-values range between 0.85 and 0.95 after repeating the same linear correlation analysis 5000 times. In these two tables it can be seen that fewer and fewer R-values are in the range between 0.85 and 0.95 with increasing disturbance. The smallest percentage is 61.0\%. If we expand the range to 0.8$\sim$0.95, the percentage of R for the case of the largest disturbance will rise to 83.1\%. All the distributions of slope K are normal, and their mean values(${\mu}_{K}$) are in the range between 1.05 and 1.16. Their standard deviation(${\sigma}_{K}$) are less than 0.08.

In light of the above analysis, we suggest that the effect of the inaccurate redshift is small for the results obtained in Sect. 4.1 and Sect. 4.2. Even if a sample contains a small amount of BL Lacs without measured redshift, our main results will not change. The correlation between $log {\nu}_{peak}^{syn}$ and -1/$c_{syn}$ for BL Lacs can still be explained by stochastical particle acceleration mechanisms. In addition, the IC component of BL Lacs can also be considered as an SSC process. However, this conclusion only applies to our sub-sample. It is possible that, if the sample size or the disturbance of redshift increase, our results will be completely different.

\begin{table*}
\begin{minipage}{155mm}
\centering
\caption{The distribution of R, p and K in the correlation between $log {\nu}_{peak}^{syn}$ and -1/$c_{syn}$ for all BL Lacs in sub-sample A.}
\begin{tabular}{@{}llrrr@{}}
\hline\hline
${\mu}_{disturbance}$   	&	${\mu}_{R}$	&	The percentage of $p<0.0001$	&	${\mu}_{K}$	\\
	&	${\sigma}_{R}$	&		&	${\sigma}_{K}$	\\
\hline
50\%z	&	0.700	&	100\%	&	1.87	\\
	&	0.01	&		&	0.01	\\
40\%z	&	0.700	&	100\%	&	1.87	\\
	&	0.004	&		&	0.01	\\
30\%z	&	0.701	&	100\%	&	1.87	\\
	&	0.003	&		&	0.01	\\
20\%z	&	0.702	&	100\%	&	1.87	\\
	&	0.002	&		&	0.005	\\
10\%z	&	0.702	&	100\%	&	1.87	\\
	&	0.001	&		&	0.002	\\
\hline
\end{tabular}
\end{minipage}
\end{table*}

\begin{table*}
\begin{minipage}{155mm}
\centering
\caption{The distribution of R, p and K in the correlation between $log {\nu}_{peak}^{syn}$ and -1/$c_{syn}$ for BL Lacs with measured redshift in sub-sample A.}
\begin{tabular}{@{}llrrr@{}}
\hline\hline
${\mu}_{disturbance}$   	&	${\mu}_{R}$	&	The percentage of $p<0.0001$	&	${\mu}_{K}$	\\
	&	${\sigma}_{R}$	&		&	${\sigma}_{K}$	\\
\hline
50\%z	&	0.65	&	100\%	&	1.89	\\
	&	0.02	&		&	0.04	\\
40\%z	&	0.66	&	100\%	&	1.89	\\
	&	0.01	&		&	0.03	\\
30\%z	&	0.67	&	100\%	&	1.89	\\
	&	0.01	&		&	0.02	\\
20\%z	&	0.67	&	100\%	&	1.89	\\
	&	0.01	&		&	0.01	\\
10\%z	&	0.66	&	100\%	&	1.89	\\
	&	0.003	&		&	0.01	\\
\hline
\end{tabular}
\end{minipage}
\end{table*}

\begin{table*}
\begin{minipage}{155mm}
\centering
\caption{The distribution of R, p and K in the correlation between $log L_{IC}$ and $log L_{syn}$ for all BL Lacs in sub-sample B.}
\begin{tabular}{@{}llrr@{}}
\hline\hline
${\mu}_{disturbance}$  	&	the percentage of R $\in$(0.85, 0.95) 	&	the percentage of $p<0.0001$	&	${\mu}_{K}$	\\
	&		&		&	${\sigma}_{K}$	\\
\hline
50\%z	&	70.0\%	&	99.7\%	&	1.05 	\\
	&		&		&	0.07 	\\
40\%z	&	74.9\%	&	99.9\%	&	1.06 	\\
	&		&		&	0.06 	\\
30\%z	&	82.2\%	&	100\%	&	1.07 	\\
	&		&		&	0.05 	\\
20\%z	&	92.9\%	&	100\%	&	1.09 	\\
	&		&		&	0.04 	\\
10\%z	&	98.5\%	&	100\%	&	1.12 	\\
	&		&		&	0.03	\\
\hline
\end{tabular}
\end{minipage}
\end{table*}

\begin{table*}
\begin{minipage}{155mm}
\centering
\caption{The distribution of R, p and K in the correlation between $log L_{IC}$ and $log L_{syn}$ for BL Lacs with measured redshift in sub-sample B.}
\begin{tabular}{@{}llrr@{}}
\hline\hline
${\mu}_{disturbance}$  	&	the percentage of R $\in$(0.85, 0.95) 	&	the percentage of $p<0.0001$	&	${\mu}_{K}$	\\
	&		&		&	${\sigma}_{K}$	\\
\hline
50\%z	&	61.0\%	&	95.1\%	&	1.09 	\\
	&		&		&	0.08 	\\
40\%z	&	66.0\%	&	97.3\%	&	1.10 	\\
	&		&		&	0.08 	\\
30\%z	&	77.7\%	&	99.3\%	&	1.11 	\\
	&		&		&	0.07 	\\
20\%z	&	90.9\%	&	100\%	&	1.13 	\\
	&		&		&	0.06 	\\
10\%z	&	99.5\%	&	100\%	&	1.16 	\\
	&		&		&	0.04	\\
\hline
\end{tabular}
\end{minipage}
\end{table*}

\subsection{Particle acceleration mechanisms for FSRQ and BL Lac}
The correlation between $log {\nu}_{peak}^{syn}$ and -1/$c_{syn}$ can be explained in two different scenarios, namely the statistical(energy-dependent acceleration probability and fluctuation of fractional acceleration gain) and the stochastical acceleration mechanisms.

FSRQ and BL Lac, the two different blazar subclasses have many differences. These include(Giommi et al. 2012b): (i) different optical spectral; (ii) different extended radio powers; (iii) very different redshift distributions; (iv) different cosmological evolutions; (v) widely different mix of FSRQs and BL Lac objects in radio and X-ray selected samples; (vi) widely different distributions of the synchrotron peak energy ${\nu}_{peak}^{S}$. Therefore, in this paper, we collect a large sample and separate them into FSRQs(N=200) and BL Lacs(N=79) in order to study the linear correlations between  $log {\nu}_{peak}^{syn}$ and -1/$c_{syn}$, respectively. For BL Lacs, the slope of the correlation($k_{BL Lac}$=1.87$\pm$0.19) is consistent with the stochastical acceleration mechanisms. However, for FSRQs, the slope of the correlation($k_{FSRQ}$=3.69$\pm$0.24) has a big difference. It is not consistent with any theoretical values(k=5/2, 10/3 and 2) and cannot be explained by the two particle acceleration mechanisms(Chen 2014).

Chen (2014) used a sample of 43 blazars in order to study the linear correlation between $log {\nu}_{peak}^{syn}$ and -1/$c_{syn}$. The slope of the correlation was 2.04$\pm$0.03, which is consistent with the stochastic acceleration mechanisms. The number of objects in the sample was too small to separate them into FSRQs and BL Lacs. Morever, the correlation was based on only eight HSP blazars. Perhaps this is why he did not find different slopes between FSRQs and BL Lacs.

As noted above, the slope of BL Lacs is consistent with the results of Chen (2014) and can be explained by stochastic particle accelerations. For FSRQs, the slope of the correlation $k_{FSRQ}$=3.69$\pm$0.24 is close to 10/3, which can be explained by statistical particle acceleration for the case of fluctuation of fractional acceleration gain. Tramacere et al. (2011) pointed out that the statistic method does not give a complete physical description of the processes responsible for the systematic and stochastic energy gain, as it ignores various physical processes. Thus the slope of the correlation that we found for FSRQs implies that some physical processes may cause error in statistic method. Even so, from the perspective of particle acceleration mechanisms, FSRQs are different from BL Lacs. As far as we know, in terms of the particle acceleration mechanisms, this is the first time that a difference between FSRQs and BL Lacs has been found by using a large sample. Our results can provide an observational information that is relevant to particle acceleration models.

\subsection{EC vs SSC and Curvature of FSRQ}
A typical SED of blazars displays two peaks. The high-energy component is usually explained as arising from IC scattering of the same electrons as produce the synchrotron emission. Depending on the origin of the soft photons, the IC scattering can be divided into EC and SSC components. According to Ghisellini et al. (1996), there are two relations that can determine whether the high-energy component is dominated by EC or SSC($L_{EC}$$\sim$$L_{syn}^{1.5}$, $L_{SSC}$$\sim$$L_{syn}^{1.0}$). Here we study the correlation between $log L_{IC}$ and $log L_{syn}$ for FSRQs and BL Lacs in our subsamples. For FSRQs, the slope of the best linear regression is 1.45$\pm$0.11; this suggests that the high-energy component of FSRQs is dominated by the EC process(Ghisellini et al. 2002; Celotti $\&$ Ghisellini 2008; Ghisellini et al. 2010; Finke 2013). The result of BL Lacs($k_{BL Lac}$=1.12$\pm$0.10) suggest that the high-energy component is dominated by the SSC process(Zhang et al. 2013; Lister et al. 2011; Celotti $\&$ Ghisellini 2008; Ghisellini et al. 2002; Ghisellini et al. 2010; Ackermann et al. 2012).

When characterizing the two components of the SEDs of blazars, the curvature is another important parameter. It can represent the value of bolomeric flux/luminosity if the peak frequency and peak flux/luminosity are known. By studying the correlation between $log L_{BLR}$ and -1/c, we found that there is no correlation for synchrotron component but that the IC component shows a significant correlation. The synchrotron component is generally explained by synchrotron emission from relativistic electrons in a jet(Maraschi, Ghisellini $\&$ Celotti 1992; Massaro et al. 2006; Hovatta et al. 2009), which means that there is no correlation between $log L_{BLR}$ and -1/$c_{syn}$. The origin of soft photons in IC scattering is complex. In one of the EC models, the soft photons could come from the BLR(Sikora et al. 1994; Dermer et al. 1997). The significant correlation between $log L_{BLR}$ and -1/$c_{IC}$ that we found might suggest that the soft photons of IC scattering are indeed mainly from the BLR.

In addition, we studied the correlation between -1/c and the $log L_{bol}$, $log M_{BH}$ and $\Gamma$. Coincidentally, all these parameters have a significant correlation with -1/$c_{IC}$  and no correlation with -1/$c_{syn}$. Perhaps there is a deep-seated physical signification, or maybe it is just a statistical coincidence.

\subsection{Jet of FSRQ}

In current theoretical models of the formation of jet, power is generated through accretion and the extraction of rotational energy of disc/black hole(Blandford $\&$ Znajek 1977; Blandford $\&$ Payne 1982) and is then converted into the kinetic power of jet.

Bolometric luminosity is one of the most important parameters of blazars and can function as an index of the jet power(Du et al. 2016). The broad-line luminosity can be taken as an indicator of accretion power(Celotti, Padovani $\&$ Ghisellini 1997). Black holes will be spun up through accretion, as these objects acquire mass and angular momentum simultaneously through accretion (Chai, Cao $\&$ Gu 2012). The presence of the jet implies that the gravitational potential energy of the falling matter not only can be transformed into heat and radiation, but can also amplify the magnetic field, allowing the field to access the large store of black hole rotational energy and transform part of it into the mechanical power of the jet, as discusses by Ghisellini et al. (2014). These authors predicted that jet power is depended on $(aMB)^2$, where a and M are respectively the spin and mass of the black hole and B is the magnetic field at its horizon.

In our work, the correlations between $log L_{bol}$ and $M_{BH}$, $L_{BLR}$ are significant. Our results show that the jet is correlated both with black hole and the accretion disc, which suggest that the origin of jet is a mixture of the mechanisms proposed by Blandford $\&$ Znajek and by Blandford $\&$ Payne(Punsly $\&$ Coroniti 1990; Meier et al. 1999, 2001). Furthermore, the coefficient of the $L_{BLR}$$\sim$$L_{bol}$ correlation is 0.9032, very close to 1. This result is consistent with Xie et al. (2007)(($log L_{BLR}$=$log L_{jet}$+$log {\eta}$+const)). The correlation between $log L_{bol}$ and $\Gamma$ can reflect the $P_{power}^{jet}$$\sim$$\Gamma$ relationship. Our result shows that the correlation between $log L_{bol}$ and $\Gamma$ is significant, and thus the correlation between $P_{power}^{jet}$ and $\Gamma$ is significant, too. From the results concerning $L_{bol}$$\sim$$\Gamma$ and $L_{bol}$$\sim$$M_{BH}$, we found that more powerful jets, the plasma blob in jets will move faster(Kharb et al. 2010; Wu et al. 2011; L$\ddot{u}$ et al. 2012) and have larger black hole masses(Xiong et al. 2014a). In other words, FSRQs with a larger black hole mass have faster jets($M_{BH}$$\sim$$\Gamma$). According to Maraschi $\&$ Tavecchio 2003($\delta$$\sim$$\Gamma$), the Doppler factor($\delta$) is also the indicator of the jet speed, so $M_{BH}$$\sim$$\delta$ can be used to represent $M_{BH}$$\sim$$\Gamma$. In Arshakian et al. (2005), the significant correlation between $M_{BH}$ and $\delta$ was found using a sample of 12 objects. Torrealba et al. (2008) obtained the same result using the 15 objects.

\section*{Acknowledgments}

We thank the anonymous referee for insightful comments and constructive suggestions. Part of this work is based on archival data, software or online services provided by the ASI SCIENCE DATA CENTER (ASDC). We are very grateful to professor Giommi and supporting team in ASDC for their help. This work is supported by the Joint Research Fund in Astronomy (Grant Nos U1431123, 11263006, 11463001) under cooperative agreement between the National Natural Science Foundation of China (NSFC) and Chinese Academy of Sciences (CAS), the Provincial Natural Science Foundation of Yunnan (Grant No.: 2013FZ042) and the Yunnan province education department project (Grant No. 2014Y138).

\begin{table*}
\tiny
 \centering
  \caption{The sample.}
  \begin{tabular}{@{}llrrrrlrlrlr@{}}
  \hline\hline
   Fermi name  &   Z  &   time    &   $Log{\nu}_{syn}^{peak}$    &   $c_{syn}$    &  $logM_{BH}$  & $logL_{BLR}$ & $\Gamma$ & $LogL_{syn}$ & $logL_{bol}$ &  Blazar type. \\
    &  &  &  &   $c_{IC}$  & Ref. & Ref. & Ref. & $LogL_{IC}$ &  &  \\
(1) &   (2) &   (3) &   (4) &   (5) &   (6) &   (7) &   (8) &   (9) &  (10) & (11) \\
 \hline
2FGL J0004.7-4736	&	0.88	&	2010/5/27-2010/6/4      	&	13.22 	&	-0.17 	&	7.85	&	-	&	11.77	&	46.78 	&	-	&	FSRQ-LSP       	\\
	&		&		&		&	-	&	1	&	-	&	2	&	-	&		&		\\
2FGL J0006.1+3821	&	0.229	&	2010/5/27-2010/6/4      	&	13.05 	&	-0.26 	&	-	&	-	&	-	&	45.76 	&	-	&	FSRQ-LSP       	\\
	&		&		&		&	-	&	-	&	-	&	-	&	-	&		&		\\
2FGL J0011.3+0054	&	1.4934	&	2010/6/23	&	13.25 	&	-0.20 	&	7.8,7.09 	&	-	&	13	&	47.02 	&	-	&	FSRQ-LSP       	\\
	&		&		&		&	-	&	3	&	-	&	3	&	-	&		&		\\
2FGL J0017.4-0018	&	1.574	&	2010/6/24	&	12.95 	&	-0.12 	&	8.55, 9.04	&	-	&	14.53	&	46.57 	&	47.90 	&	FSRQ-LSP       	\\
	&		&		&		&	-0.11 	&	1	&	-	&	2	&	47.88 	&		&		\\
2FGL J0017.6-0510	&	0.226	&	2010/6/22	&	13.29 	&	-0.19 	&	7.55	&	-	&	6.53	&	45.27 	&	-	&	FSRQ-LSP       	\\
	&		&		&		&	-	&	1	&	-	&	2	&	-	&		&		\\
2FGL J0023.2+4454	&	1.062	&	2010/1/12	&	13.49 	&	-0.14 	&	7.78	&	-	&	11.98	&	46.38 	&	-	&	FSRQ-LSP       	\\
	&		&		&		&	-	&	1	&	-	&	2	&	-	&		&		\\
2FGL J0030.2-4223	&	0.495	&	2010/6/7-2010/6/8       	&	13.55 	&	-0.13 	&	-	&	-	&	-	&	46.01 	&	-	&	FSRQ-LSP       	\\
	&		&		&		&	-	&	-	&	-	&	-	&	-	&		&		\\
2FGL J0038.3-2457	&	1.196	&	2010/6/19	&	12.77 	&	-0.25 	&	-	&	-	&	-	&	46.98 	&	-	&	FSRQ-LSP       	\\
	&		&		&		&	-	&	-	&	-	&	-	&	-	&		&		\\
2FGL J0043.7+3426	&	0.966	&	2010/1/12	&	-	&	-	&	8.01	&	44.02	&	13	&	-	&	-	&	FSRQ  	\\
	&		&		&		&	-0.04 	&	2	&	2	&	2	&	47.06 	&		&		\\
2FGL J0046.7-8416	&	1.032	&	2010/4/2-2010/4/6       	&	13.15 	&	-0.17 	&	8.68	&	-	&	12.52	&	46.61 	&	-	&	FSRQ-LSP       	\\
	&		&		&		&	-	&	1	&	-	&	2	&	-	&		&		\\
2FGL J0047.9+2232	&	1.161	&	2010/1/8	&	13.96 	&	-0.12 	&	8.43, 8.25	&	-	&	14.67	&	46.16 	&	-	&	FSRQ-LSP       	\\
	&		&		&		&	-	&	1	&	-	&	2	&	-	&		&		\\
2FGL J0049.7-5738	&	1.797	&	2010/5/26-2010/6/10     	&	12.95 	&	-0.15 	&	-	&	-	&	-	&	47.38 	&	47.85 	&	FSRQ-LSP       	\\
	&		&		&		&	-0.09 	&	-	&	-	&	-	&	47.67 	&		&		\\
2FGL J0050.1-0452	&	0.922	&	2010/6/30	&	13.22 	&	-0.16 	&	8.2	&	-	&	10.61	&	46.24 	&	-	&	FSRQ-LSP       	\\
	&		&		&		&	-	&	1	&	-	&	2	&	-	&		&		\\
2FGL J0051.0-0648	&	1.975	&	2009/5/17	&	12.62 	&	-0.21 	&	-	&	-	&	-	&	47.33 	&	48.25 	&	FSRQ-LSP       	\\
	&		&		&		&	-0.07 	&	-	&	-	&	-	&	48.20 	&		&		\\
2FGL J0057.9+3311	&	1.369	&	2010/1/13-2010/1/14     	&	13.10 	&	-0.20 	&	8.01, 7.97	&	44.21	&	13.42	&	46.53 	&	-	&	FSRQ-LSP       	\\
	&		&		&		&	-	&	2	&	2	&	2	&	-	&		&		\\
2FGL J0102.3+4216	&	0.874	&	2010/1/17-2010/1/19     	&	13.05 	&	-0.25 	&	7.92, 7.49	&	-	&	12	&	46.40 	&	-	&	FSRQ-LSP       	\\
	&		&		&		&	-	&	3	&	-	&	3	&	-	&		&		\\
2FGL J0102.7+5827	&	0.644	&	2010/1/23	&	12.93 	&	-0.20 	&	7.57	&	-	&	11	&	46.83 	&	-	&	FSRQ-LSP       	\\
	&		&		&		&	-	&	1	&	-	&	3	&	-	&		&		\\
2FGL J0105.0-2411	&	1.747	&	2010/6/25-2010/7/4      	&	12.77 	&	-0.29 	&	8.85, 8.97	&	-	&	15.98	&	47.24 	&	-	&	FSRQ-LSP       	\\
	&		&		&		&	-	&	1	&	-	&	2	&	-	&		&		\\
2FGL J0108.6+0135	&	2.107	&	2010/7/10	&	13.02 	&	-0.17 	&	8.83	&	46.13	&	28.47	&	47.89 	&	-	&	FSRQ-LSP       	\\
	&		&		&		&	-	&	2	&	2	&	2	&	-	&		&		\\
2FGL J0109.9+6132	&	0.785	&	2010/1/31-2010/2/3      	&	12.73 	&	-0.28 	&	-	&	-	&	-	&	46.97 	&	48.15 	&	FSRQ-LSP       	\\
	&		&		&		&	-0.14 	&	-	&	-	&	-	&	48.12 	&		&		\\
2FGL J0112.8+3208	&	0.603	&	2009/8/31	&	0.00 	&	-	&	-	&	-	&	-	&	-	&	-	&	FSRQ  	\\
	&		&		&		&	-0.08 	&	-	&	-	&	-	&	47.40 	&		&		\\
2FGL J0113.7+4948	&	0.395	&	2010/8/27	&	13.39 	&	-0.15 	&	8.34	&	-	&	7.34	&	46.08 	&	46.44 	&	FSRQ-LSP       	\\
	&		&		&		&	-0.07 	&	1	&	-	&	2	&	46.18 	&		&		\\
2FGL J0116.0-1134	&	0.671	&	2010/6/7	&	13.49 	&	-0.12 	&	8.57,8.92	&	-	&	-	&	46.65 	&	-	&	FSRQ-LSP       	\\
	&		&		&		&	-	&	10	&	-	&	-	&	-	&		&		\\
2FGL J0118.8-2142	&	1.165	&	2010/6/29-2010/7/4      	&	13.17 	&	-0.19 	&	-	&	-	&	-	&	47.10 	&	-	&	FSRQ-LSP       	\\
	&		&		&		&	-	&	-	&	-	&	-	&	-	&		&		\\
2FGL J0132.8-1654	&	1.02	&	2010/7/7-2010/8/4       	&	13.64 	&	-0.14 	&	-	&	-	&	-	&	47.24 	&	-	&	FSRQ-LSP       	\\
	&		&		&		&	-	&	-	&	-	&	-	&	-	&		&		\\
2FGLJ0136.9+4751 	&	0.859	&	2010/2/5	&	12.84 	&	-0.18 	&	8.73, 8.3	&	-	&	-	&	47.30 	&	47.99 	&	FSRQ-LSP       	\\
	&		&		&		&	-0.09 	&	-	&	-	&	-	&	47.90 	&		&		\\
2FGL J0137.6-2430	&	0.835	&	2010/7/2-2010/7/6       	&	13.25 	&	-0.15 	&	9.11, 9.13	&	-	&	11.87	&	46.89 	&	-	&	FSRQ-LSP       	\\
	&		&		&		&	-	&	1	&	-	&	2	&	-	&		&		\\
2FGL J0145.1-2732	&	1.148	&	2010/7/2-2010/7/6       	&	13.09 	&	-0.16 	&	-	&	-	&	-	&	46.96 	&	-	&	FSRQ-LSP       	\\
	&		&		&		&	-	&	-	&	-	&	-	&	-	&		&		\\
2FGL J0158.0-4609	&	2.287	&	2010/5/2	&	12.92 	&	-0.18 	&	7.98, 8.52	&	-	&	16.85	&	46.76 	&	47.88 	&	FSRQ-LSP       	\\
	&		&		&		&	-0.09 	&	1	&	-	&	2	&	47.84 	&		&		\\
2FGL J0205.3-1657	&	1.466	&	2010/1/8	&	12.85 	&	-0.17 	&	-	&	-	&	-	&	47.05 	&	47.98 	&	FSRQ-LSP       	\\
	&		&		&		&	-0.09 	&	-	&	-	&	-	&	47.93 	&		&		\\
2FGL J0206.5-1149	&	1.663	&	2010/1/12-2010/1/13     	&	13.74 	&	-0.16 	&	-	&	-	&	-	&	47.23 	&	-	&	FSRQ-LSP       	\\
	&		&		&		&	-	&	-	&	-	&	-	&	-	&		&		\\
2FGL J0217.7+7353	&	2.367	&	2010/9/10-2010/9/11     	&	12.74 	&	-0.21 	&	-	&	-	&	-	&	47.81 	&	50.07 	&	FSRQ-LSP       	\\
	&		&		&		&	-0.11 	&	-	&	-	&	-	&	50.07 	&		&		\\
2FGL J0217.5-0813	&	0.607	&	2010/3/5	&	12.84 	&	-0.22 	&	6.53	&	-	&	7.8	&	46.14 	&	-	&	FSRQ-LSP       	\\
	&		&		&		&	-	&	1	&	-	&	2	&	-	&		&		\\
2FGLJ0217.9+0143 	&	1.715	&	2010/1/29-2010/2/2      	&	13.22 	&	-0.18 	&	-	&	-	&	-	&	47.89 	&	48.26 	&	FSRQ-LSP   	\\
	&		&		&		&	-0.17 	&	-	&	-	&	-	&	48.01 	&		&		\\
2FGL J0221.0+3555	&	0.944	&	2010/8/4-2010/8/20      	&	13.00 	&	-0.24 	&	-	&	-	&	-	&	46.93 	&	47.72 	&	FSRQ-LSP       	\\
	&		&		&		&	-0.07 	&	-	&	-	&	-	&	47.65 	&		&		\\
2FGL J0222.0-1615	&	0.7	&	2010/5/30	&	13.32 	&	-0.14 	&	-	&	-	&	-	&	46.11 	&	47.08 	&	FSRQ-LSP       	\\
	&		&		&		&	-0.08 	&	-	&	-	&	-	&	47.03 	&		&		\\
2FGL J0229.3-3644	&	2.115	&	2010/6/7	&	12.98 	&	-0.23 	&	-	&	-	&	-	&	47.21 	&	-	&	FSRQ-LSP       	\\
	&		&		&		&	-	&	-	&	-	&	-	&	-	&		&		\\
2FGL J0230.8+4031	&	1.019	&	2010/6/3	&	13.07 	&	-0.18 	&	-	&	-	&	-	&	46.64 	&	-	&	FSRQ-LSP       	\\
	&		&		&		&	-	&	-	&	-	&	-	&	-	&		&		\\
2FGL J0237.1-6136	&	0.467	&	2010/6/12-2010/6/16     	&	13.56 	&	-0.18 	&	-	&	-	&	-	&	46.51 	&	48.49 	&	FSRQ-LSP       	\\
	&		&		&		&	-0.07 	&	-	&	-	&	-	&	48.49 	&		&		\\
2FGLJ0237.8+2846 	&	1.213	&	2010/2/5	&	13.29 	&	-0.17 	&	9.22	&	45.9	&	12.86	&	47.62 	&	48.42 	&	FSRQ-LSP   	\\
	&		&		&		&	-0.15 	&	1	&	13	&	2	&	48.34 	&		&		\\
2FGL J0245.1+2406	&	2.247	&	2010/02/01-2010/3/11    	&	13.66 	&	-0.10 	&	9.12, 9.18	&	-	&	20.88	&	46.87 	&	48.94 	&	FSRQ-LSP       	\\
	&		&		&		&	-0.10 	&	1	&	-	&	2	&	48.93 	&		&		\\
2FGL J0245.9-4652	&	1.385	&	2010/7/3-2010/7/8       	&	13.68 	&	-0.13 	&	8.48, 8.32	&	-	&	20.21	&	47.32 	&	-	&	FSRQ-LSP       	\\
	&		&		&		&	-	&	1	&	-	&	2	&	-	&		&		\\
2FGL J0250.6+1713	&	1.1	&	2010/1/30	&	13.82 	&	-0.15 	&	-	&	-	&	-	&	46.06 	&	-	&	FSRQ-LSP       	\\
	&		&		&		&	-	&	-	&	-	&	-	&	-	&		&		\\
2FGL J0252.7-2218	&	1.419	&	2010/6/7	&	13.08 	&	-0.18 	&	9.4	&	-	&	19.47	&	46.92 	&	-	&	FSRQ-LSP       	\\
	&		&		&		&	-	&	1	&	-	&	2	&	-	&		&		\\
2FGL J0253.5+5107	&	1.732	&	2010/2/10	&	12.77 	&	-0.19 	&	9.11,7.37	&	-	&	14	&	46.64 	&	-	&	FSRQ-LSP       	\\
	&		&		&		&	-	&	3	&	-	&	3	&	-	&		&		\\
2FGL J0257.7-1213	&	1.391	&	2010/1/22	&	13.52 	&	-0.14 	&	9.22	&	45.14	&	13.2	&	46.64 	&	-	&	FSRQ-LSP       	\\
	&		&		&		&	-	&	2	&	2	&	2	&	-	&		&		\\
2FGL J0259.5+0740	&	0.893	&	2010/1/29	&	13.07 	&	-0.15 	&	-	&	-	&	-	&	46.46 	&	-	&	FSRQ-LSP       	\\
	&		&		&		&	-	&	-	&	-	&	-	&	-	&		&		\\
2FGL J0302.7-7919	&	1.053	&	2010/4/15	&	13.10 	&	-0.20 	&	-	&	-	&	-	&	46.52 	&	-	&	FSRQ-LSP       	\\
	&		&		&		&	-	&	-	&	-	&	-	&	-	&		&		\\
2FGL J0303.5-6209	&	1.348	&	2010/6/12	&	12.92 	&	-0.18 	&	9.76	&	-	&	15.43	&	47.23 	&	-	&	FSRQ-LSP       	\\
	&		&		&		&	-	&	1	&	-	&	2	&	-	&		&		\\
2FGL J0310.0-6058	&	1.479	&	2010/6/19	&	13.11 	&	-0.17 	&	8.87	&	44.88	&	16.68	&	47.25 	&	-	&	FSRQ-LSP       	\\
	&		&		&		&	-	&	2	&	2	&	2	&	-	&		&		\\
2FGL J0310.7+3813	&	0.816	&	2010/2/9	&	12.74 	&	-0.22 	&	8.23	&	43.82	&	9.76	&	46.34 	&	-	&	FSRQ-LSP       	\\
	&		&		&		&	-	&	2	&	2	&	2	&	-	&		&		\\

\hline
\end{tabular}
\end{table*}

\addtocounter{table}{-1}
\begin{table*}
\tiny
 \centering
  \caption{$Continued.$}
  \begin{tabular}{@{}llrrrrlrlrlr@{}}
  \hline\hline
  Fermi name  &   Z  &   time    &   $Log{\nu}_{syn}^{peak}$    &   $c_{syn}$    &  $logM_{BH}$  & $logL_{BLR}$ & $\Gamma$ & $LogL_{syn}$ & $logL_{bol}$ &  Blazar type. \\
    &  &  &  &   $c_{IC}$  & Ref. & Ref. & Ref. & $LogL_{IC}$ &  &  \\
(1) &   (2) &   (3) &   (4) &   (5) &   (6) &   (7) &   (8) &   (9) &  (10) & (11) \\
 \hline
2FGL J0315.8-1024	&	1.565	&	2010/1/28	&	13.05 	&	-0.17 	&	7.17, 8.33	&	44.67	&	7.23	&	46.38 	&	-	&	FSRQ-LSP       	\\
	&		&		&		&	-	&	2	&	2	&	2	&	-	&		&		\\
2FGL J0326.1+2226	&	2.066	&	2010/2/8	&	13.43 	&	-0.16 	&	9.5, 9.16	&	45.81	&	20.67	&	47.77 	&	-	&	FSRQ-LSP       	\\
	&		&		&		&	-	&	2	&	2	&	2	&	-	&		&		\\
2FGLJ0337.0+3200 	&	1.259	&	2010/1/17	&	-	&	-	&	9.25	&	45.93	&	27.48	&	-	&	-	&	FSRQ  	\\
	&		&		&		&	-0.06 	&	4	&	4	&	8	&	48.37 	&		&		\\
2FGL J0342.4+3859	&	0.945	&	2010/2/15-2010/2/16     	&	13.65 	&	-0.17 	&	7.42	&	43.87	&	12.11	&	46.28 	&	-	&	FSRQ-LSP       	\\
	&		&		&		&	-	&	2	&	2	&	2	&	-	&		&		\\
2FGLJ0350.0-2104 	&	2.944	&	2008/10/15	&	-	&	-	&	-	&	-	&	-	&	-	&	-	&	FSRQ  	\\
	&		&		&		&	-0.11 	&	-	&	-	&	-	&	48.74 	&		&		\\
2FGL J0405.8-1309	&	0.571	&	2010/2/9-2010/2/10      	&	13.47 	&	-0.11 	&	9.08,9.07	&	45.25	&	8.59	&	46.49 	&	-	&	FSRQ-LSP       	\\
	&		&		&		&	-	&	2	&	2	&	2	&	-	&		&		\\
2FGL J0407.7+0740	&	1.133	&	2010/2/14-2010/2/15     	&	12.99 	&	-0.22 	&	8.65	&	44.51	&	12.73	&	46.78 	&	47.49 	&	FSRQ-LSP       	\\
	&		&		&		&	-0.09 	&	2	&	2	&	2	&	47.39 	&		&		\\
2FGL J0413.5-5332	&	1.024	&	2009/10/17	&	-	&	-	&	7.83	&	44.14	&	13.08	&	-	&	-	&	FSRQ  	\\
	&		&		&		&	-0.07 	&	2	&	2	&	2	&	47.41 	&		&		\\
2FGL J0422.1-0645	&	0.242	&	2010/2/14	&	14.12 	&	-0.13 	&	7.47	&	43.42	&	5.79	&	45.52 	&	-	&	FSRQ-ISP       	\\
	&		&		&		&	-	&	2	&	2	&	2	&	-	&		&		\\
2FGL J0423.2-0120 	&	0.916	&	2009/8/27	&	13.41 	&	-0.14 	&	9.03, 9, 8.41	&	44.63	&	11.2	&	47.84 	&	48.22 	&	 FSRQ-LSP  	\\
	&		&		&		&	-0.08 	&	1	&	8	&	2	&	47.98 	&		&		\\
2FGL J0439.0-1252	&	1.285	&	2010/7/1	&	13.07 	&	-0.18 	&	8.66	&	44.78	&	12.11	&	46.75 	&	47.47 	&	FSRQ-LSP       	\\
	&		&		&		&	-0.07 	&	2	&	2	&	2	&	47.38 	&		&		\\
2FGLJ0457.0-2325 	&	1.003	&	2010/2/25	&	13.20 	&	-0.16 	&	8.8	&	-	&	-	&	47.18 	&	47.90 	&	FSRQ-LSP   	\\
	&		&		&		&	-0.11 	&	7	&	-	&	-	&	47.81 	&		&		\\
2FGL J0501.2-0155	&	2.291	&	2010/2/27	&	13.53 	&	-0.12 	&	9.27, 8.66	&	45.3	&	16.57	&	47.79 	&	-	&	FSRQ-LSP       	\\
	&		&		&		&	-	&	2	&	2	&	2	&	-	&		&		\\
2FGL J0507.5-6102	&	1.089	&	2010/1/20	&	13.54 	&	-0.12 	&	8.74	&	44.86	&	13.62	&	46.58 	&	-	&	FSRQ-LSP       	\\
	&		&		&		&	-	&	2	&	2	&	2	&	-	&		&		\\
2FGL J0516.5-4601	&	0.194	&	2010/2/20	&	13.23 	&	-0.13 	&	8.02	&	-	&	4.86	&	45.26 	&	-	&	FSRQ-LSP       	\\
	&		&		&		&	-	&	2	&	2	&	2	&	-	&		&		\\
2FGLJ0530.8+1333 	&	2.07	&	2009/9/24	&	-	&	-	&	10.2, 9.4	&	-	&	21.35	&	-	&	-	&	FSRQ  	\\
	&		&		&		&	-0.14 	&	1	&	-	&	2	&	48.86 	&		&		\\
2FGL J0532.7+0733	&	1.254	&	2010/4/25	&	12.98 	&	-0.16 	&	8.43	&	44.86	&	17.76	&	47.10 	&	48.12 	&	FSRQ-LSP       	\\
	&		&		&		&	-0.09 	&	2	&	2	&	2	&	48.07 	&		&		\\
2FGLJ0539.3-2841 	&	3.104	&	2010/3/12	&	12.87 	&	-0.23 	&	-	&	-	&	-	&	47.76 	&	49.00 	&	FSRQ-LSP   	\\
	&		&		&		&	-0.21 	&	-	&	-	&	-	&	48.98 	&		&		\\
2FGL J0601.1-7037	&	2.409	&	2010/3/9	&	13.58 	&	-0.19 	&	7.36	&	44.69	&	22.49	&	47.71 	&	48.56 	&	FSRQ-LSP       	\\
	&		&		&		&	-0.10 	&	2	&	2	&	2	&	48.49 	&		&		\\
2FGL J0608.0-0836	&	0.872	&	2010/6/7	&	12.62 	&	-0.16 	&	8.43, 8.825	&	44.97	&	22.62	&	46.63 	&	47.28 	&	FSRQ-LSP       	\\
	&		&		&		&	-0.16 	&	2	&	2	&	2	&	47.17 	&		&		\\
2FGL J0608.0-1521	&	1.094	&	2010/3/16-2010/3/23     	&	13.24 	&	-0.18 	&	8.09	&	44.51	&	14.77	&	46.38 	&	-	&	FSRQ-LSP       	\\
	&		&		&		&	-	&	2	&	2	&	2	&	-	&		&		\\
2FGL J0635.5-7516	&	0.653	&	2009/1/13	&	-	&	-	&	9.41, 8.81	&	45.23	&	12.43	&	-	&	-	&	FSRQ  	\\
	&		&		&		&	-0.08 	&	2	&	2	&	2	&	47.34 	&		&		\\
2FGL J0654.2+4514	&	0.928	&	2010/3/23	&	13.13 	&	-0.19 	&	8.17	&	44.26	&	14.94	&	46.64 	&	47.59 	&	FSRQ-LSP       	\\
	&		&		&		&	-0.09 	&	2	&	2	&	2	&	47.54 	&		&		\\
2FGL J0654.5+5043	&	1.253	&	2010/1/15	&	14.01 	&	-0.14 	&	7.86, 8.79	&	43.97	&	6.3	&	-	&	-	&	FSRQ-ISP       	\\
	&		&		&		&	-0.05 	&	2	&	2	&	2	&	47.59 	&		&		\\
2FGL J0656.2-0320	&	0.634	&	2010/3/31	&	12.58 	&	-0.22 	&	8.82, 8.77	&	45.68	&	12.08	&	46.22 	&	47.07 	&	FSRQ-LSP       	\\
	&		&		&		&	-0.21 	&	2	&	2	&	2	&	47.01 	&		&		\\
2FGL J0714.0+1933	&	0.54	&	2010/4/2-2010/4/3       	&	13.51 	&	-0.19 	&	7.33, 7.91	&	43.93	&	11.3	&	46.43 	&	47.09 	&	FSRQ-LSP       	\\
	&		&		&		&	-0.10 	&	2	&	2	&	2	&	46.98 	&		&		\\
2FGL J0721.5+0404	&	0.665	&	2010/4/5	&	12.92 	&	-0.20 	&	8.49, 9.12	&	45.33	&	9.41	&	46.25 	&	-	&	FSRQ-LSP       	\\
	&		&		&		&	-	&	2	&	2	&	2	&	-	&		&		\\
2FGL J0739.2+0138	&	0.189	&	2010/4/10-2010/4/11     	&	13.76 	&	-0.13 	&	8, 8.47, 7.86	&	44.19	&	16.57	&	45.89 	&	-	&	FSRQ-LSP       	\\
	&		&		&		&	-	&	2	&	2	&	2	&	-	&		&		\\
2FGL J0746.6+2549	&	2.979	&	2010/10/15	&	12.78 	&	-0.21 	&	9.59, 9.23	&	45.46	&	25.46	&	47.27 	&	48.68 	&	FSRQ-LSP       	\\
	&		&		&		&	-0.11 	&	2	&	2	&	2	&	48.67 	&		&		\\
2FGL J0750.6+1230	&	0.889	&	210/4/1-2010/4/12       	&	12.95 	&	-0.19 	&	8.15	&	44.95	&	12.82	&	47.15 	&	47.53 	&	FSRQ-LSP       	\\
	&		&		&		&	-0.07 	&	2	&	2	&	2	&	47.30 	&		&		\\
2FGL J0805.5+6145	&	3.033	&	2010/4/3-2010/4/4       	&	12.79 	&	-0.17 	&	9.07	&	45.56	&	26.84	&	47.40 	&	48.57 	&	FSRQ-LSP       	\\
	&		&		&		&	-0.11 	&	2	&	2	&	2	&	48.54 	&		&		\\
2FGL J0824.7+3914	&	1.216	&	2010/4/13	&	13.05 	&	-0.13 	&	8.55	&	44.83	&	12.78	&	46.86 	&	-	&	FSRQ-LSP       	\\
	&		&		&		&	-	&	2	&	2	&	2	&	-	&		&		\\
2FGLJ0824.9+5552 	&	1.417	&	2010/3/28	&	-	&	-	&	9.42, 9.1	&	-	&	16.34	&	-	&	-	&	FSRQ  	\\
	&		&		&		&	-0.14 	&	2	&	2	&	2	&	48.17 	&		&		\\
2FGL J0834.3+4221	&	0.249	&	2010/4/13-2010/4/15     	&	13.67 	&	-0.14 	&	9.68	&	43.07	&	6.52	&	45.58 	&	-	&	FSRQ-LSP       	\\
	&		&		&		&	-	&	2	&	2	&	2	&	-	&		&		\\
2FGL J0841.6+7052 	&	2.218	&	2010/3/21	&	-	&	-	&	9.36	&	-	&	28.15	&	-	&	-	&	FSRQ  	\\
	&		&		&		&	-0.14 	&	1	&	-	&	2	&	48.30 	&		&		\\
2FGL J0903.4+4651	&	1.466	&	2010/4/18	&	12.90 	&	-0.12 	&	9.25	&	45.26	&	12.72	&	46.93 	&	-	&	FSRQ-LSP       	\\
	&		&		&		&	-	&	2	&	2	&	2	&	-	&		&		\\
2FGL J0909.1+0121	&	1.024	&	2010/4/18-2010/5/5      	&	13.44 	&	-0.15 	&	9.32, 8.55, 9.14	&	45.2	&	16.57	&	47.17 	&	48.23 	&	FSRQ-LSP       	\\
	&		&		&		&	-0.12 	&	2	&	2	&	2	&	48.19 	&		&		\\
2FGL J0910.9+2246	&	2.661	&	2010/4/28	&	13.11 	&	-0.22 	&	8.7	&	45.21	&	20.46	&	47.14 	&	-	&	FSRQ-LSP       	\\
	&		&		&		&	-	&	2	&	2	&	2	&	-	&		&		\\
2FGL J0912.1+4126	&	2.563	&	2010/2/25	&	12.97 	&	-0.31 	&	9.32	&	45.36	&	17.42	&	47.30 	&	47.96 	&	FSRQ-LSP       	\\
	&		&		&		&	-0.07 	&	2	&	2	&	2	&	47.85 	&		&		\\
2FGL J0917.0+3900	&	1.267	&	2010/4/25	&	12.84 	&	-0.16 	&	8.62	&	44.8	&	6.52	&	46.64 	&	-	&	FSRQ-LSP       	\\
	&		&		&		&	-	&	2	&	2	&	2	&	-	&		&		\\
2FGL J0920.9+4441	&	2.19	&	2009/10/29	&	12.97 	&	-0.20 	&	9.25, 9.31, 9.29	&	45.775	&	25.67	&	47.83 	&	48.88 	&	FSRQ-LSP       	\\
	&		&		&		&	-0.09 	&	2	&	2	&	2	&	48.83 	&		&		\\
2FGL J0923.2+4125	&	1.732	&	2010/4/25-2010/4/26     	&	13.67 	&	-0.13 	&	7.68, 8.16	&	43.75	&	16.29	&	46.86 	&	-	&	FSRQ-LSP       	\\
	&		&		&		&	-	&	2	&	2	&	2	&	-	&		&		\\
2FGL J0924.0+2819	&	0.744	&	2010/4/29	&	12.88 	&	-0.19 	&	8.8,8.825	&	44.52	&	10.47	&	46.32 	&	-	&	FSRQ-LSP       	\\
	&		&		&		&	-	&	2	&	2	&	2	&	-	&		&		\\
2FGL J0937.6+5009	&	0.275	&	2010/4/21-2010/4/25     	&	12.85 	&	-0.26 	&	8.29, 7.5	&	44.26	&	5.86	&	45.60 	&	-	&	FSRQ-LSP       	\\
	&		&		&		&	-	&	2	&	2	&	2	&	-	&		&		\\
2FGL J0948.8+4040	&	1.249	&	2010/4/30	&	12.68 	&	-0.17 	&	8.95	&	45.5	&	30.86	&	46.88 	&	-	&	FSRQ-LSP       	\\
	&		&		&		&	-	&	2	&	2	&	2	&	-	&		&		\\
2FGL J0956.9+2516	&	0.707	&	2010/5/7-2010/6/15      	&	12.88 	&	-0.19 	&	9.34, 9,8.7, 8.465	&	44.93	&	11.33	&	46.60 	&	47.22 	&	FSRQ-LSP       	\\
	&		&		&		&	-0.08 	&	2	&	2	&	2	&	47.10 	&		&		\\
2FGL J0957.7+5522	&	0.896	&	2009/11/1	&	13.82 	&	-0.10 	&	8.96, 7.87, 8.07, 8.45	&	44.58	&	17.16	&	46.92 	&	47.94 	&	FSRQ-LSP       	\\
	&		&		&		&	-0.05 	&	2	&	2	&	2	&	47.90 	&		&		\\
2FGL J1012.6+2440	&	1.805	&	2010/5/11	&	14.02 	&	-0.10 	&	7.73, 7.86	&	44.56	&	19.64	&	46.42 	&	-	&	FSRQ-ISP       	\\
	&		&		&		&	-	&	2	&	2	&	2	&	-	&		&		\\
2FGL J1014.1+2306	&	0.566	&	2010/5/12	&	13.48 	&	-0.12 	&	8.479, 8.54	&	44.89	&	9	&	46.23 	&	-	&	FSRQ-LSP       	\\
	&		&		&		&	-	&	2	&	2	&	2	&	-	&		&		\\
2FGL J1016.0+0513	&	1.713	&	2010/5/18	&	12.91 	&	-0.19 	&	9.11, 7.99	&	44.62	&	20.18	&	46.87 	&	-	&	FSRQ-LSP       	\\
	&		&		&		&	-	&	2	&	2	&	2	&	-	&		&		\\
2FGL J1017.0+3531	&	1.228	&	2010/5/8	&	13.04 	&	-0.15 	&	9.1	&	45.34	&	13.16	&	46.64 	&	-	&	FSRQ-LSP       	\\
	&		&		&		&	-	&	2	&	2	&	2	&	-	&		&		\\
2FGL J1033.2+4117	&	1.117	&	2010/5/8-2010/5/9       	&	12.93 	&	-0.20 	&	8.65, 8.61	&	44.48	&	13.94	&	47.25 	&	-	&	FSRQ-LSP       	\\
	&		&		&		&	-	&	2	&	2	&	2	&	-	&		&		\\

\hline
\end{tabular}
\end{table*}

\addtocounter{table}{-1}
\begin{table*}
\tiny
 \centering
  \caption{$Continued.$}
  \begin{tabular}{@{}llrrrrlrlrlr@{}}
  \hline\hline
  Fermi name  &   Z  &   time    &   $Log{\nu}_{syn}^{peak}$    &   $c_{syn}$    &  $logM_{BH}$  & $logL_{BLR}$ & $\Gamma$ & $LogL_{syn}$ & $logL_{bol}$ &  Blazar type. \\
    &  &  &  &   $c_{IC}$  & Ref. & Ref. & Ref. & $LogL_{IC}$ &  &  \\
(1) &   (2) &   (3) &   (4) &   (5) &   (6) &   (7) &   (8) &   (9) &  (10) & (11) \\
 \hline
2FGL J1037.5-2820	&	1.066	&	2010/1/22-2010/1/23     	&	13.62 	&	-0.14 	&	8.99	&	44.95	&	14.34	&	46.65 	&	47.54 	&	FSRQ-LSP       	\\
	&		&		&		&	-0.11 	&	2	&	2	&	2	&	47.48 	&		&		\\
2FGL J1106.1+2814	&	0.843	&	2010/5/24	&	13.46 	&	-0.18 	&	8.85	&	45.16	&	10.7	&	46.67 	&	47.28 	&	FSRQ-LSP       	\\
	&		&		&		&	-0.10 	&	2	&	2	&	2	&	47.16 	&		&		\\
2FGL J1112.4+3450	&	1.949	&	2010/5/19-2010/5/25	&	13.25 	&	-0.21 	&	9.04, 8.78	&	45.22	&	19.45	&	47.34 	&	-	&	FSRQ-LSP       	\\
	&		&		&		&	-	&	2	&	2	&	2	&	-	&		&		\\
2FGL J1120.4+0710	&	1.336	&	2010/6/5	&	13.57 	&	-0.14 	&	8.83	&	44.47	&	12.85	&	46.57 	&	-	&	FSRQ-LSP       	\\
	&		&		&		&	-	&	2	&	2	&	2	&	-	&		&		\\
2FGL J1124.2+2338	&	1.549	&	2010/5/28-2010/5/30     	&	13.01 	&	-0.19 	&	8.79	&	45.05	&	14.44	&	47.00 	&	-	&	FSRQ-LSP       	\\
	&		&		&		&	-	&	2	&	2	&	2	&	-	&		&		\\
2FGLJ1126.6-1856 	&	1.048	&	2010/6/10	&	13.58 	&	-0.13 	&	-	&	-	&	-	&	47.30 	&	47.92 	&	FSRQ-LSP   	\\
	&		&		&		&	-0.09 	&	-	&	-	&	-	&	47.80 	&		&		\\
2FGLJ1130.3-1448 	&	1.184	&	2009/12/28	&	-	&	-	&	9.18	&	-	&	16.96	&	-	&	-	&	FSRQ  	\\
	&		&		&		&	-0.13 	&	1	&	-	&	2	&	47.99 	&		&		\\
2FGL J1146.8-3812	&	1.048	&	2010/6/24	&	12.97 	&	-0.20 	&	8.5	&	44.48	&	13.78	&	47.20 	&	47.68 	&	FSRQ-LSP       	\\
	&		&		&		&	-0.06 	&	2	&	2	&	2	&	47.51 	&		&		\\
2FGL J1146.9+4000	&	1.089	&	2010/5/25-2010/5/26     	&	13.42 	&	-0.16 	&	8.98, 8.93	&	45.06	&	14.94	&	47.15 	&	-	&	FSRQ-LSP       	\\
	&		&		&		&	-	&	2	&	2	&	2	&	-	&		&		\\
2FGL J1152.4-0840	&	2.367	&	2010/6/16	&	13.18 	&	-0.17 	&	9.38	&	45.25	&	18.7	&	47.60 	&	-	&	FSRQ-LSP       	\\
	&		&		&		&	-	&	2	&	2	&	2	&	-	&		&		\\
2FGL J1159.5+2914	&	0.724	&	2010/5/28-2010/6/11     	&	13.49 	&	-0.13 	&	9.18, 7.9, 8.54, 8.375	&	44.68	&	25.9	&	47.01 	&	47.72 	&	FSRQ-LSP       	\\
	&		&		&		&	-0.08 	&	2	&	2	&	2	&	47.63 	&		&		\\
2FGL J1206.0-2638	&	0.789	&	2010/6/30-2010/12/10    	&	13.43 	&	-0.12 	&	8.59, 9	&	44.07	&	12	&	46.51 	&	47.20 	&	FSRQ-LSP       	\\
	&		&		&		&	-0.08 	&	2	&	2	&	2	&	47.10 	&		&		\\
2FGL J1208.8+5441	&	1.344	&	2010/5/16-2010/5/21     	&	13.71 	&	-0.13 	&	8.67, 8.4	&	44.52	&	16.95	&	46.83 	&	48.03 	&	FSRQ-LSP       	\\
	&		&		&		&	-0.12 	&	2	&	2	&	2	&	48.00 	&		&		\\
2FGL J1209.7+1807	&	0.845	&	2010/6/9-2010/6/11      	&	12.88 	&	-0.25 	&	8.94, 8.515	&	44.47	&	9.86	&	46.18 	&	-	&	FSRQ-LSP       	\\
	&		&		&		&	-	&	2	&	2	&	2	&	-	&		&		\\
2FGL J1219.7+0201	&	0.241	&	2010/6/24	&	-	&	-	&	8.87	&	-	&	6.22	&	-	&	-	&	FSRQ  	\\
	&		&		&		&	-0.08 	&	1	&	-	&	2	&	46.15 	&		&		\\
2FGLJ1222.4+0413 	&	0.967	&	2010/6/17-2010/6/29     	&	13.57 	&	-0.13 	&	8.24, 8.37	&	-	&	14.94	&	47.18 	&	48.14 	&	FSRQ-LSP   	\\
	&		&		&		&	-0.15 	&	1	&	-	&	2	&	48.09 	&		&		\\
2FGL J1228.6+4857	&	1.722	&	2009/2/25-2009/2/26     	&	-	&	-	&	9.22, 8.255	&	44.73	&	15.71	&	-	&	-	&	FSRQ  	\\
	&		&		&		&	-0.06 	&	2	&	2	&	2	&	47.93 	&		&		\\
2FGL J1239.5+0443	&	1.761	&	2009/1/2-2009/1/4       	&	-	&	-	&	8.67, 8.57	&	44.9	&	21.93	&	-	&	-	&	FSRQ  	\\
	&		&		&		&	-0.07 	&	2	&	2	&	2	&	48.33 	&		&		\\
2FGLJ1246.7-2546 	&	0.635	&	2010/1/25	&	12.92 	&	-0.25 	&	9.04	&	-	&	14.77	&	46.77 	&	47.63 	&	FSRQ-LSP   	\\
	&		&		&		&	-0.09 	&	1	&	-	&	2	&	47.57 	&		&		\\
2FGLJ1256.1-0547 	&	0.536	&	2010/1/15	&	12.62 	&	-0.18 	&	8.9, 8.43, 8.4, 8.28	&	44.78	&	20.87	&	47.27 	&	47.96 	&	FSRQ-LSP   	\\
	&		&		&		&	-0.08 	&	1	&	9	&	2	&	47.86 	&		&		\\
2FGL J1258.2+3231	&	0.806	&	2010/6/13-2010/6/19     	&	13.94 	&	-0.13 	&	8.74, 8.255	&	44.42	&	10.09	&	46.92 	&	-	&	FSRQ-LSP       	\\
	&		&		&		&	-	&	2	&	2	&	2	&	-	&		&		\\
2FGL J1303.5-4622	&	1.664	&	2010/1/24-2010/1/27     	&	13.70 	&	-0.17 	&	7.95, 8.21	&	44.21	&	14.18	&	47.06 	&	-	&	FSRQ-LSP       	\\
	&		&		&		&	-	&	2	&	2	&	2	&	-	&		&		\\
2FGLJ1310.6+3222 	&	0.997	&	2009/12/12-2009/12/21   	&	12.86 	&	-0.14 	&	8.8, 9.24, 7.3, 8.57	&	44.96	&	22.1	&	47.16 	&	47.89 	&	FSRQ-LSP       	\\
	&		&		&		&	-0.08 	&	1	&	1	&	2	&	47.79 	&		&		\\
2FGL J1317.9+3426	&	1.056	&	2010/6/16-2010/6/20     	&	13.38 	&	-0.16 	&	9.29, 9.14	&	45.08	&	10.82	&	46.74 	&	-	&	FSRQ-LSP       	\\
	&		&		&		&	-	&	2	&	2	&	2	&	-	&		&		\\
2FGL J1321.1+2215	&	0.943	&	2010/1/7	&	13.16 	&	-0.19 	&	8.42, 8.315	&	44.71	&	12.33	&	46.68 	&	-	&	FSRQ-LSP       	\\
	&		&		&		&	-	&	2	&	2	&	2	&	-	&		&		\\
2FGL J1326.8+2210	&	1.4	&	2010/3/30	&	12.88 	&	-0.19 	&	9.24,9.25	&	44.96	&	17.16	&	47.10 	&	48.03 	&	FSRQ-LSP       	\\
	&		&		&		&	-0.10 	&	2	&	2	&	2	&	47.98 	&		&		\\
2FGL J1332.5-1255	&	1.492	&	2010/7/10	&	12.75 	&	-0.24 	&	8.96, 8.61	&	45.26	&	19.63	&	46.61 	&	-	&	FSRQ-LSP       	\\
	&		&		&		&	-	&	2	&	2	&	2	&	-	&		&		\\
2FGL J1332.7+4725	&	0.668	&	2010/6/10-2010/6/19     	&	12.83 	&	-0.20 	&	8.56, 7.975	&	44.32	&	8.49	&	45.94 	&	-	&	FSRQ-LSP       	\\
	&		&		&		&	-	&	2	&	2	&	2	&	-	&		&		\\
2FGL J1333.5+5058	&	1.362	&	2010/3/13	&	13.79 	&	-0.14 	&	7.95	&	44.37	&	15.32	&	46.33 	&	47.53 	&	FSRQ-LSP       	\\
	&		&		&		&	-0.10 	&	2	&	2	&	2	&	47.50 	&		&		\\
2FGL J1337.7-1257	&	0.539	&	2010/1/18-2010/1/26     	&	12.94 	&	-0.18 	&	7.98	&	44.18	&	10.82	&	46.80 	&	47.22 	&	FSRQ-LSP       	\\
	&		&		&		&	-0.06 	&	2	&	2	&	2	&	47.00 	&		&		\\
2FGL J1344.2-1723	&	2.506	&	2010/1/21	&	13.46 	&	-0.20 	&	9.12	&	45.02	&	22.81	&	47.82 	&	-	&	FSRQ-LSP       	\\
	&		&		&		&	-	&	2	&	2	&	2	&	-	&		&		\\
2FGL J1345.4+4453	&	2.534	&	2010/6/15-2010/6/19     	&	13.08 	&	-0.16 	&	8.98	&	45.12	&	24.47	&	46.95 	&	-	&	FSRQ-LSP       	\\
	&		&		&		&	-	&	2	&	2	&	2	&	-	&		&		\\
2FGL J1345.9+0706	&	1.093	&	2010/7/5	&	13.85 	&	-0.15 	&	8.48	&	44.61	&	11.19	&	46.72 	&	-	&	FSRQ-LSP       	\\
	&		&		&		&	-	&	2	&	2	&	2	&	-	&		&		\\
2FGL J1347.7-3752	&	1.3	&	2010/9/20	&	13.77 	&	-0.15 	&	7.95, 8.62	&	44.67	&	13.71	&	46.86 	&	47.80 	&	FSRQ-LSP       	\\
	&		&		&		&	-0.10 	&	2	&	2	&	2	&	47.74 	&		&		\\
2FGL J1358.1+7644	&	1.585	&	2010/5/24-2010/5/25     	&	12.73 	&	-0.19 	&	8.34, 8.17	&	44.2	&	14.87	&	46.87 	&	47.70 	&	FSRQ-LSP       	\\
	&		&		&		&	-0.08 	&	2	&	2	&	2	&	47.64 	&		&		\\
2FGL J1359.4+5541	&	1.014	&	2010/6/6-2010/6/8       	&	13.24 	&	-0.22 	&	8	&	43.99	&	12.58	&	46.60 	&	-	&	FSRQ-LSP       	\\
	&		&		&		&	-	&	2	&	2	&	2	&	-	&		&		\\
2FGL J1408.8-0751	&	1.494	&	2010/5/23	&	13.29 	&	-0.16 	&	9.4	&	45.47	&	17.42	&	47.33 	&	48.09 	&	FSRQ-LSP       	\\
	&		&		&		&	-0.08 	&	2	&	2	&	2	&	48.00 	&		&		\\
2FGL J1419.4+3820	&	1.831	&	2010/1/8	&	12.91 	&	-0.20 	&	8.59, 8.68	&	45.1	&	16.38	&	47.32 	&	-	&	FSRQ-LSP       	\\
	&		&		&		&	-	&	2	&	2	&	2	&	-	&		&		\\
2FGL J1436.9+2319	&	1.548	&	2010/6/14	&	13.06 	&	-0.15 	&	8.44, 8.31	&	44.72	&	13.48	&	46.95 	&	47.59 	&	FSRQ-LSP       	\\
	&		&		&		&	-0.07 	&	2	&	2	&	2	&	47.48 	&		&		\\
2FGL J1504.3+1029	&	1.839	&	2010/7/29	&	13.33 	&	-0.14 	&	9.64, 8.74, 8.94	&	45.24	&	15.11	&	47.63 	&	49.21 	&	FSRQ-LSP       	\\
	&		&		&		&	-0.11 	&	2	&	2	&	2	&	49.19 	&		&		\\
2FGLJ1510.9-0545 	&	1.191	&	2010/2/13-2010/2/19     	&	13.15 	&	-0.15 	&	8.97	&	-	&	16.2	&	47.05 	&	-	&	FSRQ-LSP   	\\
	&		&		&		&	-	&	2	&	2	&	2	&	-	&		&		\\
2FGLJ1512.8-0906 	&	0.36	&	2009/1/16	&	-	&	-	&	8.6, 8.65, 8, 8.2	&	44.65	&	27.925	&	-	&	-	&	FSRQ  	\\
	&		&		&		&	-0.12 	&	4	&	4	&	4	&	48.00 	&		&		\\
2FGL J1514.6+4449	&	0.57	&	2010/4/6	&	13.05 	&	-0.27 	&	7.72, 7.62	&	43.33	&	8.66	&	46.06 	&	46.66 	&	FSRQ-LSP       	\\
	&		&		&		&	-0.09 	&	2	&	2	&	2	&	46.54 	&		&		\\
2FGL J1522.0+4348	&	2.171	&	2010/1/18	&	12.79 	&	-0.24 	&	8.59, 8.67	&	45.49	&	19.39	&	47.14 	&	-	&	FSRQ-LSP       	\\
	&		&		&		&	-	&	2	&	2	&	2	&	-	&		&		\\
2FGL J1522.1+3144	&	1.484	&	2010/1/25-2010/1/28     	&	13.43 	&	-0.11 	&	8.92	&	44.9	&	25.82	&	46.57 	&	-	&	FSRQ-LSP       	\\
	&		&		&		&	-	&	2	&	2	&	2	&	-	&		&		\\
2FGL J1539.5+2747	&	2.191	&	2010/3/17	&	13.77 	&	-0.13 	&	8.43, 8.51	&	44.63	&	15.33	&	47.15 	&	47.92 	&	FSRQ-LSP       	\\
	&		&		&		&	-0.04 	&	2	&	2	&	2	&	47.84 	&		&		\\
2FGL J1549.5+0237	&	0.414	&	2010/2/13-2010/2/20     	&	13.04 	&	-0.18 	&	8.61, 8.72, 8.47, 8.67	&	44.8	&	9.64	&	46.25 	&	46.86 	&	FSRQ-LSP       	\\
	&		&		&		&	-0.09 	&	2	&	2	&	2	&	46.74 	&		&		\\
2FGL J1553.5+1255	&	1.29	&	2010/2/11	&	12.93 	&	-0.15 	&	9.1, 8.64	&	45.19	&	17.76	&	46.74 	&	-	&	FSRQ-LSP       	\\
	&		&		&		&	-	&	2	&	2	&	2	&	-	&		&		\\
2FGL J1608.5+1029	&	1.226	&	2010/2/16-2010/2/17     	&	13.60 	&	-0.12 	&	8.64, 9.5, 8.77	&	45.04	&	18.81	&	47.23 	&	-	&	FSRQ-LSP       	\\
	&		&		&		&	-	&	2	&	2	&	2	&	-	&		&		\\
2FGL J1613.4+3409	&	1.397	&	2010/6/7	&	13.89 	&	-0.11 	&	9.12, 9.57, 9.6, 9.08	&	45.61	&	8.02	&	47.64 	&	-	&	FSRQ-LSP       	\\
	&		&		&		&	-	&	2	&	2	&	2	&	-	&		&		\\

\hline
\end{tabular}
\end{table*}

\addtocounter{table}{-1}
\begin{table*}
\tiny
 \centering
  \caption{$Continued.$}
  \begin{tabular}{@{}llrrrrlrlrlr@{}}
  \hline\hline
  Fermi name  &   Z  &   time    &   $Log{\nu}_{syn}^{peak}$    &   $c_{syn}$    &  $logM_{BH}$  & $logL_{BLR}$ & $\Gamma$ & $LogL_{syn}$ & $logL_{bol}$ &  Blazar type. \\
    &  &  &  &   $c_{IC}$  & Ref. & Ref. & Ref. & $LogL_{IC}$ &  &  \\
(1) &   (2) &   (3) &   (4) &   (5) &   (6) &   (7) &   (8) &   (9) &  (10) & (11) \\
 \hline
2FGLJ1625.7-2526 	&	0.786	&	2010/8/27-2010/9/3      	&	13.05 	&	-0.14 	&	-	&	-	&	-	&	46.92 	&	-	&	FSRQ-LSP       	\\
	&		&		&		&	-	&	-	&	-	&	-	&	-	&		&		\\
2FGLJ1635.2+3810 	&	1.814	&	2010/3/7	&	12.82 	&	-0.19 	&	9.53, 9.2, 9.67, 9.075	&	45.48	&	31.22	&	47.96 	&	49.17 	&	FSRQ-LSP   	\\
	&		&		&		&	-0.15 	&	1	&	8	&	2	&	49.14 	&		&		\\
2FGL J1637.7+4714	&	0.735	&	2010/7/30-2010/7/31     	&	12.85 	&	-0.21 	&	8.61, 8.52	&	44.58	&	12.43	&	46.60 	&	47.36 	&	FSRQ-LSP       	\\
	&		&		&		&	-0.08 	&	2	&	2	&	2	&	47.27 	&		&		\\
2FGLJ1640.7+3945 	&	1.66	&	2010/8/7	&	13.18 	&	-0.17 	&	-	&	-	&	-	&	47.38 	&	-	&	FSRQ-LSP   	\\
	&		&		&		&	-0.11 	&	-	&	-	&	-	&	-	&		&		\\
2FGLJ1642.9+3949 	&	0.593	&	2010/3/6	&	-	&	-	&	7.73,9.03	&	45.23	&	11	&	-	&	-	&	FSRQ  	\\
	&		&		&		&	-0.08 	&	3	&	8	&	3	&	47.42 	&		&		\\
2FGL J1703.2-6217	&	1.747	&	2010/3/12-2010/3/13     	&	13.23 	&	-0.23 	&	8.65, 8.55	&	45.31	&	21.47	&	47.86 	&	-	&	FSRQ-LSP       	\\
	&		&		&		&	-	&	2	&	2	&	2	&	-	&		&		\\
2FGL J1709.7+4319	&	1.027	&	2009/12/1	&	13.39 	&	-0.19 	&	7.92	&	44.03	&	14.44	&	46.75 	&	47.80 	&	FSRQ-LSP       	\\
	&		&		&		&	-0.09 	&	2	&	2	&	2	&	47.76 	&		&		\\
2FGL J1728.2+0429	&	0.296	&	2010/3/12-2010/3/13     	&	13.43 	&	-0.14 	&	8.07, 7.72	&	44.07	&	8.02	&	45.89 	&	-	&	FSRQ-LSP       	\\
	&		&		&		&	-	&	2	&	2	&	2	&	-	&		&		\\
2FGL J1733.1-1307	&	0.902	&	2010/3/14-2010/4/10     	&	13.45 	&	-0.12 	&	9.3	&	44.83	&	65.24	&	47.42 	&	48.01 	&	FSRQ-LSP       	\\
	&		&		&		&	-0.09 	&	2	&	2	&	2	&	47.88 	&		&		\\
2FGL J1740.2+5212	&	1.375	&	2010/3/6-2010/3/11      	&	13.63 	&	-0.14 	&	9.32	&	45.16	&	17.97	&	47.51 	&	-	&	FSRQ-LSP       	\\
	&		&		&		&	-	&	2	&	2	&	2	&	-	&		&		\\
2FGL J1818.6+0903	&	0.354	&	2010/3/25-2010/3/26     	&	13.71 	&	-0.17 	&	7.3, 7.5	&	43.93	&	8.22	&	45.67 	&	-	&	FSRQ-LSP       	\\
	&		&		&		&	-	&	2	&	2	&	2	&	-	&		&		\\
2FGL J1830.1+0617	&	0.745	&	2009/5/20	&	-	&	-	&	8.69, 8.86	&	45.45	&	12.52	&	-	&	-	&	FSRQ  	\\
	&		&		&		&	-0.08 	&	2	&	2	&	2	&	47.56 	&		&		\\
2FGLJ1833.6-2104 	&	2.507	&	2010/9/23	&	12.74 	&	-0.15 	&	-	&	-	&	-	&	48.06 	&	-	&	FSRQ-LSP   	\\
	&		&		&		&	-	&	-	&	-	&	-	&	-	&		&		\\
2FGL J1848.5+3216	&	0.798	&	2010/10/6-2010/10/19    	&	13.35 	&	-0.17 	&	7.87,8.21	&	44.58	&	12.06	&	46.76 	&	47.54 	&	FSRQ-LSP       	\\
	&		&		&		&	-0.07 	&	2	&	2	&	2	&	47.47 	&		&		\\
2FGL J1849.4+6706	&	0.657	&	2010/6/3-2010/7/9       	&	13.58 	&	-0.16 	&	9.14	&	44.42	&	14.27	&	47.03 	&	-	&	FSRQ-LSP       	\\
	&		&		&		&	-	&	2	&	2	&	2	&	-	&		&		\\
2FGL J1902.5-6746	&	0.254	&	2010/3/28-2010/4/6      	&	13.00 	&	-0.19 	&	7.51	&	43.35	&	6.13	&	45.30 	&	-	&	FSRQ-LSP       	\\
	&		&		&		&	-	&	2	&	2	&	2	&	-	&		&		\\
2FGLJ1911.1-2005 	&	1.119	&	2009/10/4	&	12.88 	&	-0.22 	&	-	&	-	&	-	&	47.46 	&	48.23 	&	FSRQ-LSP   	\\
	&		&		&		&	-0.11 	&	-	&	-	&	-	&	48.15 	&		&		\\
2FGLJ1923.5-2105 	&	0.874	&	2010/9/30	&	14.04 	&	-0.10 	&	-	&	-	&	-	&	47.57 	&	-	&	FSRQ-ISP   	\\
	&		&		&		&	-	&	-	&	-	&	-	&	-	&		&		\\
2FGL J1924.8-2912	&	0.353	&	2010/9/30	&	12.73 	&	-0.20 	&	9.01, 8.38	&	44.02	&	9.3	&	46.86 	&	47.15 	&	FSRQ-LSP       	\\
	&		&		&		&	-0.07 	&	2	&	2	&	2	&	46.84 	&		&		\\
2FGL J1954.6-112 	&	0.683	&	2009/12/3	&	-	&	-	&	6.73	&	43.37	&	12.23	&	-	&	-	&	FSRQ  	\\
	&		&		&		&	-0.08 	&	2	&	2	&	2	&	47.20 	&		&		\\
2FGL J1958.2-3848	&	0.63	&	2010/4/9-2010/4/14      	&	13.02 	&	-0.17 	&	7.99, 8.63	&	44.2	&	12.16	&	46.75 	&	47.44 	&	FSRQ-LSP       	\\
	&		&		&		&	-0.09 	&	2	&	2	&	2	&	47.35 	&		&		\\
2FGL J1959.1-4245	&	2.178	&	2010/4/5-2010/4/14      	&	13.41 	&	-0.20 	&	8.55, 9.41	&	45.13	&	22.62	&	47.45 	&	48.49 	&	FSRQ-LSP       	\\
	&		&		&		&	-0.11 	&	2	&	2	&	2	&	48.45 	&		&		\\
2FGL J2035.4+1058	&	0.601	&	2010/5/3-2010/5/7       	&	12.99 	&	-0.17 	&	7.74, 8.26	&	44.17	&	11.46	&	46.19 	&	-	&	FSRQ-LSP       	\\
	&		&		&		&	-	&	2	&	2	&	2	&	-	&		&		\\
2FGLJ2056.2-4715 	&	1.491	&	2010/10/18	&	12.91 	&	-0.19 	&	9.6	&	-	&	-	&	47.62 	&	-	&	FSRQ-LSP   	\\
	&		&		&		&	-	&	7	&	-	&	-	&	-	&		&		\\
2FGL J2109.9+0807	&	1.58	&	2010/5/13-2010/5/14     	&	13.57 	&	-0.15 	&	8.82	&	45.09	&	15.26	&	46.49 	&	-	&	FSRQ-LSP       	\\
	&		&		&		&	-	&	2	&	2	&	2	&	-	&		&		\\
2FGL J2115.3+2932	&	1.514	&	2010/5/25	&	12.95 	&	-0.18 	&	8.74	&	44.78	&	16.1	&	47.09 	&	-	&	FSRQ-LSP       	\\
	&		&		&		&	-	&	2	&	2	&	2	&	-	&		&		\\
2FGL J2121.0+1901	&	2.18	&	2009/1/2	&	-	&	-	&	7.75	&	44.26	&	20.75	&	-	&	-	&	FSRQ  	\\
	&		&		&		&	-0.07 	&	2	&	2	&	2	&	48.26 	&		&		\\
2FGL J2135.6-4959	&	2.181	&	2010/4/22-2010/5/5      	&	13.05 	&	-0.16 	&	8.31, 8.4	&	45.26	&	19.93	&	46.68 	&	48.12 	&	FSRQ-LSP       	\\
	&		&		&		&	-0.09 	&	2	&	2	&	2	&	48.10 	&		&		\\
2FGL J2143.5+1743	&	0.211	&	2009/1/15	&	-	&	-	&	8.6, 8.74, 8.1	&	44.26	&	8.8	&	-	&	-	&	FSRQ  	\\
	&		&		&		&	-0.13 	&	2	&	2	&	2	&	46.60 	&		&		\\
2FGL J2144.8-3356	&	1.361	&	2009/9/22-2009/9/24     	&	13.52 	&	-0.19 	&	8.31	&	44.18	&	16.18	&	47.04 	&	47.75 	&	FSRQ-LSP       	\\
	&		&		&		&	-0.07 	&	2	&	2	&	2	&	47.65 	&		&		\\
2FGLJ2148.2+0659 	&	0.999	&	2010/11/10-2010/11/19   	&	-	&	-	&	8.87	&	45.48	&	7.93	&	-	&	-	&	FSRQ  	\\
	&		&		&		&	-0.08 	&	1	&	8	&	2	&	47.05 	&		&		\\
2FGLJ2151.5-3021 	&	2.345	&	2010/5/4-2010/5/13      	&	12.92 	&	-0.13 	&	-	&	-	&	-	&	47.49 	&	48.86 	&	FSRQ-LSP   	\\
	&		&		&		&	-0.07 	&	-	&	-	&	-	&	48.84 	&		&		\\
2FGL J2157.4+3129	&	1.488	&	2009/7/8-2009/7/12      	&	13.23 	&	-0.15 	&	8.89	&	44.74	&	18.82	&	46.96 	&	48.03 	&	FSRQ-LSP       	\\
	&		&		&		&	-0.08 	&	2	&	2	&	2	&	47.99 	&		&		\\
2FGL J2157.9-1501	&	0.672	&	2010/5/17	&	13.14 	&	-0.13 	&	7.59	&	43.68	&	10.45	&	46.70 	&	-	&	FSRQ-LSP       	\\
	&		&		&		&	-	&	2	&	2	&	2	&	-	&		&		\\
2FGL J2201.9-8335	&	1.865	&	2010/7/5-2010/7/17      	&	13.31 	&	-0.15 	&	9.02, 9.16	&	45.19	&	20.18	&	47.24 	&	48.43 	&	FSRQ-LSP       	\\
	&		&		&		&	-0.12 	&	2	&	2	&	2	&	48.40 	&		&		\\
2FGL J2211.9+2355	&	1.125	&	2009/4/15-2009/4/21     	&	12.92 	&	-0.20 	&	8.46	&	44.79	&	12.08	&	46.94 	&	47.48 	&	FSRQ-LSP       	\\
	&		&		&		&	-0.04 	&	2	&	2	&	2	&	47.34 	&		&		\\
2FGL J2219.1+1805	&	1.071	&	2010/6/3-2010/6/6       	&	12.41 	&	-0.29 	&	7.65, 7.66	&	44.07	&	10.41	&	46.31 	&	-	&	FSRQ-LSP       	\\
	&		&		&		&	-	&	2	&	2	&	2	&	-	&		&		\\
2FGL J2225.6-0454	&	1.404	&	2010/5/22-2010/5/27     	&	12.75 	&	-0.16 	&	8.81, 8.54, 7.9	&	45.6	&	14.77	&	47.68 	&	48.19 	&	FSRQ-LSP       	\\
	&		&		&		&	-0.09 	&	2	&	2	&	2	&	48.03 	&		&		\\
2FGLJ2232.4+1143 	&	1.037	&	2009/11/27-2009/11/29   	&	-	&	-	&	8.7, 8.64, 9	&	45.58	&	15.47	&	-	&	-	&	FSRQ  	\\
	&		&		&		&	-0.10 	&	1	&	8	&	2	&	48.00 	&		&		\\
2FGLJ2253.9+1609 	&	0.859	&	2009/12/4-2009/12/6     	&	12.76 	&	-0.28 	&	8.7, 9.17, 8.6, 8.83	&	45.39	&	19.47	&	48.30 	&	49.36 	&	FSRQ-LSP   	\\
	&		&		&		&	-0.13 	&	1	&	8	&	2	&	49.32 	&		&		\\
2FGL J2258.0-2759	&	0.926	&	2010/5/20-2010/5/26     	&	13.14 	&	-0.17 	&	8.92, 9.16	&	45.84	&	14.94	&	47.33 	&	47.95 	&	FSRQ-LSP       	\\
	&		&		&		&	-0.10 	&	2	&	2	&	2	&	47.83 	&		&		\\
2FGL J2322.2+3206	&	1.489	&	2009/5/20	&	13.05 	&	-0.16 	&	8.66, 8.75	&	44.71	&	15.76	&	46.69 	&	47.80 	&	FSRQ-LSP       	\\
	&		&		&		&	-0.07 	&	2	&	2	&	2	&	47.76 	&		&		\\
2FGL J2327.5+0940	&	1.841	&	2010/6/18-2010/6/29     	&	12.64 	&	-0.24 	&	8.7,9.35	&	45.2	&	21.52	&	47.22 	&	48.59 	&	FSRQ-LSP       	\\
	&		&		&		&	-0.12 	&	2	&	2	&	2	&	48.57 	&		&		\\
2FGL J2334.3+0734	&	0.401	&	2009/12/20	&	12.94 	&	-0.17 	&	8.37	&	44.93	&	8.3	&	45.90 	&	46.45 	&	FSRQ-LSP       	\\
	&		&		&		&	-0.06 	&	2	&	2	&	2	&	46.31 	&		&		\\
2FGLJ2345.0-1553 	&	0.621	&	2009/1/10	&	13.40 	&	-0.18 	&	8.16, 8.48	&	44.36	&	12.4	&	46.61 	&	47.27 	&	FSRQ-LSP       	\\
	&		&		&		&	-0.08 	&	1	&	1	&	2	&	47.16 	&		&		\\
2FGL J2347.9-1629	&	0.576	&	2009/12/04-2009/12/05   	&	13.07 	&	-0.16 	&	8.72, 8.47	&	44.36	&	10.7	&	46.65 	&	47.12 	&	FSRQ-LSP       	\\
	&		&		&		&	-0.06 	&	2	&	2	&	2	&	46.94 	&		&		\\
2FGL J2356.3+0432	&	1.248	&	2010/6/22-2010/6/23     	&	12.53 	&	-0.30 	&	8.41, 8.45	&	45.02	&	12.64	&	46.32 	&	-	&	FSRQ-LSP       	\\
	&		&		&		&	-	&	2	&	2	&	2	&	-	&		&		\\

\hline
\end{tabular}
\end{table*}

\addtocounter{table}{-1}
\begin{table*}
\tiny
 \centering
  \caption{$Continued.$}
  \begin{tabular}{@{}llrrrrlrlrlr@{}}
  \hline\hline
  Fermi name  &   Z  &   time    &   $Log{\nu}_{syn}^{peak}$    &   $c_{syn}$    &  $logM_{BH}$  & $logL_{BLR}$ & $\Gamma$ & $LogL_{syn}$ & $logL_{bol}$ &  Blazar type. \\
    &  &  &  &   $c_{IC}$  & Ref. & Ref. & Ref. & $LogL_{IC}$ &  &  \\
(1) &   (2) &   (3) &   (4) &   (5) &   (6) &   (7) &   (8) &   (9) &  (10) & (11) \\
 \hline
2FGL J0007.8+4713	&	0.28	&	2010/1/12	&	15.23 	&	-0.08 	&	-	&	-	&	-	&	45.31 	&	-	&	BL Lac-HSP 	\\
	&		&		&		&	-	&	-	&	-	&	-	&	-	&		&		\\
2FGL J0009.0+0632	&	0.27	&	2010/6/25	&	12.74 	&	-0.14 	&	-	&	-	&	-	&	44.79 	&	-	&	BL Lac-LSP 	\\
	&		&		&		&	-	&	-	&	-	&	-	&	-	&		&		\\
2FGL J0012.9-3954	&	0.27	&	2010/6/3-2010/6/6     	&	12.66 	&	-0.20 	&	-	&	-	&	-	&	45.73 	&	-	&	BL Lac-LSP 	\\
	&		&		&		&	-	&	-	&	-	&	-	&	-	&		&		\\
2FGL J0021.6-2551	&	0.27	&	2010/6/15-2010/6/16   	&	14.55 	&	-0.13 	&	-	&	-	&	-	&	45.95 	&	-	&	BL Lac-ISP     	\\
	&		&		&		&	-	&	-	&	-	&	-	&	-	&		&		\\
2FGL J0022.5+0607	&	0.27	&	2010/6/28-2010/6/29   	&	13.45 	&	-0.14 	&	-	&	-	&	-	&	45.78 	&	-	&	BL Lac-LSP 	\\
	&		&		&		&	-	&	-	&	-	&	-	&	-	&		&		\\
2FGL J0029.2-7043	&	0.27	&	2010/05/02-2010/05/06	&	12.33 	&	-0.22 	&	-	&	-	&	-	&	44.96 	&	-	&	BL Lac-LSP 	\\
	&		&		&		&	-	&	-	&	-	&	-	&	-	&		&		\\
2FGL J0035.8+5951	&	0.27	&	2010/7/10	&	17.00 	&	-0.11 	&	-	&	-	&	-	&	46.81 	&	-	&	BL Lac-HSP 	\\
	&		&		&		&	-	&	-	&	-	&	-	&	-	&		&		\\
2FGL J0038.1+0015	&	0.7395	&	2010/6/29-2010/6/30   	&	13.52 	&	-0.17 	&	-	&	-	&	-	&	46.09 	&	-	&	BL Lac-LSP 	\\
	&		&		&		&	-	&	-	&	-	&	-	&	-	&		&		\\
2FGL J0045.3+2127	&	0.27	&	2009/5/29	&	15.42 	&	-0.09 	&	-	&	-	&	-	&	46.37 	&	-	&	BL Lac-HSP 	\\
	&		&		&		&	-	&	-	&	-	&	-	&	-	&		&		\\
2FGL J0050.2+0234	&	1.44	&	2010/7/2-2010/7/6     	&	13.54 	&	-0.15 	&	-	&	-	&	-	&	47.05 	&	-	&	BL Lac-LSP 	\\
	&		&		&		&	-	&	-	&	-	&	-	&	-	&		&		\\
2FGL J0050.6-0929	&	0.635	&	2009/5/24	&	15.45 	&	-0.07 	&	-	&	-	&	-	&	47.25 	&	47.29 	&	BL Lac-HSP 	\\
	&		&		&		&	-0.04 	&	-	&	-	&	-	&	46.30 	&		&		\\
2FGL J0057.9-3236	&	1.37	&	2009/09/25-2009/09/26 	&	14.31 	&	-0.12 	&	-	&	-	&	-	&	47.30 	&	-	&	BL Lac-ISP     	\\
	&		&		&		&	-	&	-	&	-	&	-	&	-	&		&		\\
2FGL J0100.2+0746	&	0.27	&	2010/7/9	&	12.76 	&	-0.24 	&	-	&	-	&	-	&	45.22 	&	46.47 	&	BL Lac-LSP 	\\
	&		&		&		&	-0.14 	&	-	&	-	&	-	&	46.45 	&		&		\\
2FGLJ0114.7+1326 	&	0.27	&	2010/7/15	&	15.07 	&	-0.11 	&	-	&	-	&	-	&	47.55 	&	48.07 	&	BL Lac-HSP 	\\
	&		&		&		&	-0.06 	&	-	&	-	&	-	&	47.91 	&		&		\\
2FGL J0120.4-2700	&	0.559	&	2010/1/27	&	14.44 	&	-0.10 	&	9.54	&	-	&	-	&	46.67 	&	-	&	BL Lac-ISP     	\\
	&		&		&		&	-	&	1	&	-	&	-	&	-	&		&		\\
2FGL J0124.5-0621	&	0.27	&	2010/7/9-2010/7/10    	&	13.80 	&	-0.15 	&	-	&	-	&	-	&	45.38 	&	-	&	BL Lac-LSP 	\\
	&		&		&		&	-	&	-	&	-	&	-	&	-	&		&		\\
2FGL J0136.5+3905	&	0.27	&	2010/1/27	&	16.40 	&	-0.08 	&	-	&	-	&	-	&	46.75 	&	-	&	BL Lac-HSP 	\\
	&		&		&		&	-	&	-	&	-	&	-	&	-	&		&		\\
2FGL J0141.5-0928	&	0.733	&	2010/5/30-2010/6/6    	&	13.46 	&	-0.17 	&	9.84	&	-	&	-	&	46.92 	&	47.33 	&	BL Lac-LSP 	\\
	&		&		&		&	-0.04 	&	1	&	-	&	-	&	47.12 	&		&		\\
2FGL J0144.6+2704	&	0.27	&	2010/1/19	&	13.47 	&	-0.17 	&	-	&	-	&	-	&	45.82 	&	-	&	BL Lac-LSP 	\\
	&		&		&		&	-	&	-	&	-	&	-	&	-	&		&		\\
2FGL J0153.9+0823	&	0.27	&	2010/3/1	&	15.15 	&	-0.10 	&	-	&	-	&	-	&	46.37 	&	46.64 	&	BL Lac-HSP 	\\
	&		&		&		&	-0.08 	&	-	&	-	&	-	&	46.29 	&		&		\\
2FGL J0203.6+7235	&	0.27	&	2009/9/11	&	-	&	-	&	-	&	-	&	-	&	-	&	-	&	BL Lac 	\\
	&		&		&		&	-0.08 	&	-	&	-	&	-	&	46.27 	&		&		\\
2FGLJ0204.0+3045 	&	0.761	&	2010/1/25	&	12.42 	&	-0.17 	&	-	&	-	&	-	&	45.82 	&	-	&	BL Lac-LSP 	\\
	&		&		&		&	-	&	-	&	-	&	-	&	-	&		&		\\
2FGLJ0210.7-5102 	&	1.003	&	2009/11/26	&	13.30 	&	-0.15 	&	9.8	&	-	&	-	&	47.32 	&	48.08 	&	BL Lac-LSP 	\\
	&		&		&		&	-0.08 	&	7	&	-	&	-	&	47.99 	&		&		\\
2FGLJ0238.7+1637 	&	0.94	&	2010/1/30	&	13.09 	&	-0.25 	&	8	&	-	&	-	&	47.42 	&	48.31 	&	BL Lac-LSP 	\\
	&		&		&		&	-0.09 	&	4	&	-	&	-	&	48.25 	&		&		\\
2FGLJ0334.2-4008 	&	1.445	&	2010/1/17-2010/1/18   	&	13.18 	&	-0.17 	&	-	&	-	&	-	&	47.64 	&	48.18 	&	BL Lac-LSP 	\\
	&		&		&		&	-0.06 	&	-	&	-	&	-	&	48.04 	&		&		\\
2FGLJ0428.6-3756 	&	1.03	&	2010/8/17	&	13.19 	&	-0.20 	&	-	&	-	&	-	&	47.54 	&	-	&	BL Lac-LSP 	\\
	&		&		&		&	-	&	-	&	-	&	-	&	-	&		&		\\
2FGLJ0516.8-6207 	&	0.27	&	2009/1/15	&	-	&	-	&	-	&	-	&	-	&	-	&	-	&	BL Lac 	\\
	&		&		&		&	-0.06 	&	-	&	-	&	-	&	46.10 	&		&		\\
2FGLJ0523.0-3628 	&	0.055	&	2010/3/5	&	13.38 	&	-0.18 	&	8.477	&	43.38	&	5	&	45.32 	&	45.58 	&	BL Lac-LSP 	\\
	&		&		&		&	-0.07 	&	6	&	9	&	6	&	45.24 	&		&		\\
2FGLJ0538.8-4405 	&	0.892	&	2010/3/3	&	13.74 	&	-0.14 	&	8.778	&	45.31	&	13	&	48.09 	&	-	&	BL Lac-LSP 	\\
	&		&		&		&	-	&	6	&	9	&	6	&	-	&		&		\\
2FGLJ0712.9+5032 	&	0.27	&	2009/1/21	&	13.72 	&	-0.18 	&	-	&	-	&	-	&	46.39 	&	46.59 	&	BL Lac-LSP 	\\
	&		&		&		&	-0.05 	&	-	&	-	&	-	&	46.15 	&		&		\\
2FGLJ0721.9+7120 	&	0.27	&	2005/4/4	&	14.62 	&	-0.12 	&	-	&	-	&	-	&	48.29 	&	48.64 	&	BL Lac-ISP     	\\
	&		&		&		&	-0.06 	&	-	&	-	&	-	&	48.38 	&		&		\\
2FGLJ0738.0+1742 	&	0.424	&	2010/10/7	&	14.04 	&	-0.16 	&	8.2	&	-	&	-	&	46.57 	&	-	&	BL Lac-ISP     	\\
	&		&		&		&	-	&	7	&	-	&	-	&	-	&		&		\\
2FGLJ0818.2+4223 	&	0.53	&	2010/10/15	&	13.22 	&	-0.14 	&	-	&	-	&	-	&	46.46 	&	-	&	BL Lac-LSP 	\\
	&		&		&		&	-	&	-	&	-	&	-	&	-	&		&		\\
2FGLJ0854.8+2005 	&	0.306	&	2010/4/10	&	13.87 	&	-0.12 	&	8.79	&	43.6	&	14.289	&	46.64 	&	46.78 	&	BL Lac-LSP 	\\
	&		&		&		&	-0.18 	&	4	&	4	&	4	&	46.23 	&		&		\\
2FGLJ0909.2+2308 	&	0.223	&	2010/5/22-2010/5/26   	&	15.17 	&	-0.11 	&	-	&	-	&	-	&	45.31 	&	-	&	BL Lac-HSP 	\\
	&		&		&		&	-	&	-	&	-	&	-	&	-	&		&		\\
2FGLJ0915.8+2932 	&	0.101	&	2010/10/28	&	16.04 	&	-0.11 	&	-	&	-	&	-	&	45.27 	&	-	&	BL Lac-HSP 	\\
	&		&		&		&	-	&	-	&	-	&	-	&	-	&		&		\\
2FGLJ0958.6+6533 	&	0.367	&	2010/3/12	&	13.27 	&	-0.25 	&	-	&	-	&	-	&	46.59 	&	46.90 	&	BL Lac-LSP 	\\
	&		&		&		&	-0.07 	&	-	&	-	&	-	&	46.60 	&		&		\\
2FGLJ1001.0+2913 	&	0.558	&	2010/5/7-2010/5/8     	&	13.31 	&	-0.17 	&	-	&	-	&	-	&	46.45 	&	-	&	BL Lac-LSP 	\\
	&		&		&		&	-	&	-	&	-	&	-	&	-	&		&		\\
2FGLJ1015.1+4925 	&	0.212	&	2010/4/28-2010/5/1    	&	15.53 	&	-0.13 	&	-	&	-	&	-	&	46.06 	&	-	&	BL Lac-HSP 	\\
	&		&		&		&	-	&	-	&	-	&	-	&	-	&		&		\\
2FGLJ1043.1+2404 	&	0.559117	&	2010/7/9	&	13.28 	&	-0.16 	&	-	&	-	&	-	&	46.48 	&	46.80 	&	BL Lac-LSP 	\\
	&		&		&		&	-0.07 	&	-	&	-	&	-	&	46.51 	&		&		\\
2FGLJ1057.0-8004 	&	0.581	&	2010/8/30	&	12.97 	&	-0.18 	&	-	&	-	&	-	&	46.78 	&	-	&	BL Lac-LSP 	\\
	&		&		&		&	-	&	-	&	-	&	-	&	-	&		&		\\
2FGLJ1058.4+0133 	&	0.888	&	2009/12/3	&	13.14 	&	-0.14 	&	7.37	&	-	&	14	&	47.47 	&	47.83 	&	BL Lac-LSP 	\\
	&		&		&		&	-0.04 	&	3	&	-	&	3	&	47.57 	&		&		\\
2FGLJ1058.6+5628 	&	0.14333	&	2010/4/18-2010/5/4    	&	15.07 	&	-0.15 	&	-	&	-	&	-	&	45.51 	&	-	&	BL Lac-HSP 	\\
	&		&		&		&	-	&	-	&	-	&	-	&	-	&		&		\\
2FGLJ1104.4+3812 	&	0.03	&	2009/11/15-2009/11/17 	&	16.13 	&	-0.13 	&	8.22	&	41.4	&	3.3	&	45.90 	&	45.94 	&	BL Lac-HSP 	\\
	&		&		&		&	-0.20 	&	4	&	4	&	4	&	44.88 	&		&		\\
2FGLJ1217.8+3006 	&	0.13	&	2009/12/3-2009/12/19  	&	14.82 	&	-0.10 	&	-	&	-	&	-	&	45.59 	&	45.88 	&	BL Lac-ISP     	\\
	&		&		&		&	-0.07 	&	-	&	-	&	-	&	45.57 	&		&		\\
2FGLJ1136.7+7009 	&	0.046	&	2008/10/30	&	15.68 	&	-0.07 	&	-	&	-	&	-	&	44.61 	&	-	&	BL Lac-HSP 	\\
	&		&		&		&	-	&	-	&	-	&	-	&	-	&		&		\\
2FGLJ1141.9+1550 	&	0.299	&	2010/6/4-2010/6/8     	&	12.59 	&	-0.27 	&	-	&	-	&	-	&	45.19 	&	-	&	BL Lac-LSP 	\\
	&		&		&		&	-	&	-	&	-	&	-	&	-	&		&		\\
2FGLJ1146.8-3812 	&	1.048	&	2010/6/24	&	12.81 	&	-0.18 	&	8.5	&	-	&	13.78	&	46.97 	&	47.47 	&	BL Lac-LSP 	\\
	&		&		&		&	-0.04 	&	1	&	-	&	8	&	47.30 	&		&		\\
2FGLJ1204.3-0711 	&	0.184	&	2010/8/9	&	15.01 	&	-0.17 	&	-	&	-	&	-	&	45.44 	&	-	&	BL Lac-HSP 	\\
	&		&		&		&	-	&	-	&	-	&	-	&	-	&		&		\\
2FGLJ1221.3+3010 	&	0.18365	&	2010/6/7	&	16.42 	&	-0.13 	&	-	&	-	&	-	&	45.94 	&	-	&	BL Lac-HSP 	\\
	&		&		&		&	-	&	-	&	-	&	-	&	-	&		&		\\
2FGLJ1221.4+2814 	&	0.102	&	2009/12/10-2009/12/12 	&	14.48 	&	-0.15 	&	-	&	-	&	-	&	45.50 	&	45.78 	&	BL Lac-ISP     	\\
	&		&		&		&	-0.05 	&	-	&	-	&	-	&	45.45 	&		&		\\

\hline
\end{tabular}
\end{table*}

\addtocounter{table}{-1}
\begin{table*}
\tiny
 \centering
  \caption{$Continued.$}
  \begin{tabular}{@{}llrrrrlrlrlr@{}}
  \hline\hline
  Fermi name  &   Z  &   time    &   $Log{\nu}_{syn}^{peak}$    &   $c_{syn}$    &  $logM_{BH}$  & $logL_{BLR}$ & $\Gamma$ & $LogL_{syn}$ & $logL_{bol}$ &  Blazar type. \\
    &  &  &  &   $c_{IC}$  & Ref. & Ref. & Ref. & $LogL_{IC}$ &  &  \\
(1) &   (2) &   (3) &   (4) &   (5) &   (6) &   (7) &   (8) &   (9) &  (10) & (11) \\
 \hline
 2FGLJ1248.2+5820 	&	0.8474	&	2010/5/20	&	15.30 	&	-0.16 	&	-	&	-	&	-	&	47.41 	&	47.64 	&	BL Lac-HSP 	\\
	&		&		&		&	-0.09 	&	-	&	-	&	-	&	47.26 	&		&		\\
2FGLJ1427.0+2347 	&	0.16	&	2010/1/22-2010/1/23   	&	14.94 	&	-0.15 	&	-	&	-	&	-	&	46.14 	&	-	&	BL Lac-ISP     	\\
	&		&		&		&	-	&	-	&	-	&	-	&	-	&		&		\\
2FGLJ1428.6+4240 	&	0.129172	&	2010/6/28	&	17.68 	&	-0.09 	&	-	&	-	&	-	&	45.48 	&	-	&	BL Lac-HSP 	\\
	&		&		&		&	-	&	-	&	-	&	-	&	-	&		&		\\
2FGLJ1442.7+1159 	&	0.16309	&	2010/2/26-2010/3/9    	&	16.94 	&	-0.09 	&	-	&	-	&	-	&	45.36 	&	-	&	BL Lac-HSP 	\\
	&		&		&		&	-	&	-	&	-	&	-	&	-	&		&		\\
2FGLJ1443.9-3908 	&	0.065	&	2009/9/24	&	15.80 	&	-0.14 	&	-	&	-	&	-	&	44.99 	&	-	&	BL Lac-HSP 	\\
	&		&		&		&	-	&	-	&	-	&	-	&	-	&		&		\\
2FGLJ1517.7-2421 	&	0.048	&	2010/2/20	&	13.79 	&	-0.15 	&	-	&	-	&	-	&	44.93 	&	45.20 	&	BL Lac-LSP 	\\
	&		&		&		&	-0.03 	&	-	&	-	&	-	&	44.88 	&		&		\\
2FGLJ1542.9+6129 	&	0.117	&	2009/1/18-2009/1/20   	&	14.84 	&	-0.11 	&	-	&	-	&	-	&	45.19 	&	45.66 	&	BL Lac-ISP     	\\
	&		&		&		&	-0.05 	&	-	&	-	&	-	&	45.48 	&		&		\\
2FGLJ1653.9+3945 	&	0.033	&	2010/3/21	&	16.30 	&	-0.07 	&	8.62	&	41.36	&	4.2	&	44.94 	&	45.08 	&	BL Lac-HSP 	\\
	&		&		&		&	-0.15 	&	4	&	4	&	4	&	44.51 	&		&		\\
2FGLJ1719.3+1744 	&	0.137	&	2009/1/8	&	13.55 	&	-0.14 	&	-	&	-	&	-	&	45.14 	&	46.25 	&	BL Lac-LSP 	\\
	&		&		&		&	-0.03 	&	-	&	-	&	-	&	46.22 	&		&		\\
2FGLJ1728.2+5015 	&	0.055	&	2010/5/1	&	16.10 	&	-0.13 	&	-	&	-	&	-	&	44.66 	&	-	&	BL Lac-HSP 	\\
	&		&		&		&	-	&	-	&	-	&	-	&	-	&		&		\\
2FGLJ1744.1+1934 	&	0.084	&	2010/3/20-2010/3/27   	&	-	&	-	&	-	&	-	&	-	&	-	&	-	&	BL Lac 	\\
	&		&		&		&	-0.09 	&	-	&	-	&	-	&	44.54 	&		&		\\
2FGLJ1751.5+0938 	&	0.322	&	2010/4/1	&	12.68 	&	-0.21 	&	8.34	&	-	&	8.892	&	46.36 	&	46.70 	&	BL Lac-LSP 	\\
	&		&		&		&	-0.11 	&	4	&	-	&	4	&	46.43 	&		&		\\
2FGLJ1800.5+7829 	&	0.68	&	2009/10/13	&	13.88 	&	-0.12 	&	7.92	&	44.56	&	1.1	&	47.27 	&	47.65 	&	BL Lac-LSP 	\\
	&		&		&		&	-0.06 	&	4	&	4	&	4	&	47.42 	&		&		\\
2FGLJ1806.7+6948 	&	0.051	&	2009/11/3-2009/11/5   	&	14.25 	&	-0.08 	&	-	&	-	&	-	&	44.57 	&	-	&	BL Lac-ISP     	\\
	&		&		&		&	-	&	-	&	-	&	-	&	-	&		&		\\
2FGLJ2000.0+6509 	&	0.047	&	2009/9/26	&	16.43 	&	-0.09 	&	8.09	&	-	&	-	&	45.18 	&	-	&	BL Lac-HSP 	\\
	&		&		&		&	-	&	5	&	-	&	-	&	-	&		&		\\
2FGLJ2009.5-4850 	&	0.071	&	2009/10/5	&	15.60 	&	-0.09 	&	9.03	&	-	&	-	&	45.65 	&	-	&	BL Lac-HSP 	\\
	&		&		&		&	-	&	5	&	-	&	-	&	-	&		&		\\
2FGLJ2022.5+7614 	&	0.594	&	2010/6/12-2010/6/16   	&	14.41 	&	-0.08 	&	-	&	-	&	-	&	46.68 	&	-	&	BL Lac-ISP     	\\
	&		&		&		&	-	&	-	&	-	&	-	&	-	&		&		\\
2FGLJ2039.1-1046 	&	0.27	&	2010/4/28-2010/4/29   	&	13.75 	&	-0.15 	&	-	&	-	&	-	&	46.10 	&	-	&	BL Lac-LSP 	\\
	&		&		&		&	-	&	-	&	-	&	-	&	-	&		&		\\
2FGLJ2133.8-0154 	&	1.284	&	2010/5/16-2010/5/17   	&	13.09 	&	-0.19 	&	-	&	-	&	-	&	47.55 	&	-	&	BL Lac-LSP 	\\
	&		&		&		&	-	&	-	&	-	&	-	&	-	&		&		\\
2FGLJ2152.4+1735 	&	0.871	&	2010/4/8-2010/4/10    	&	13.70 	&	-0.15 	&	-	&	-	&	-	&	46.76 	&	47.14 	&	BL Lac-LSP 	\\
	&		&		&		&	-0.10 	&	-	&	-	&	-	&	46.91 	&		&		\\
2FGLJ2158.8-3013 	&	0.116	&	2010/5/12	&	15.11 	&	-0.16 	&	-	&	-	&	-	&	46.22 	&	-	&	BL Lac-HSP 	\\
	&		&		&		&	-	&	-	&	-	&	-	&	-	&		&		\\
2FGLJ2202.8+4216 	&	0.069	&	2009/12/23-2009/12/26 	&	-	&	-	&	8.23	&	42.38	&	6.99842	&	-	&	-	&	BL Lac 	\\
	&		&		&		&	-0.10 	&	4	&	4	&	4	&	44.71 	&		&		\\
2FGLJ2247.2-0002 	&	0.949	&	2010/1/14-2010/1/16   	&	13.62 	&	-0.15 	&	-	&	-	&	-	&	46.77 	&	47.17 	&	BL Lac-LSP 	\\
	&		&		&		&	-0.04 	&	-	&	-	&	-	&	46.95 	&		&		\\
2FGLJ2258.8-5524 	&	0.479	&	2010/8/5	&	15.83 	&	-0.16 	&	-	&	-	&	-	&	46.06 	&	-	&	BL Lac-HSP 	\\
	&		&		&		&	-	&	-	&	-	&	-	&	-	&		&		\\
2FGLJ2315.7-5014 	&	0.808	&	2010/5/17-2010/5/18   	&	13.35 	&	-0.16 	&	-	&	-	&	-	&	46.77 	&	-	&	BL Lac-LSP 	\\
	&		&		&		&	-	&	-	&	-	&	-	&	-	&		&		\\
2FGLJ2323.8+4212 	&	0.059	&	2010/2/20-2010/2/26   	&	15.82 	&	-0.24 	&	-	&	-	&	-	&	44.45 	&	-	&	BL Lac-HSP 	\\
	&		&		&		&	-	&	-	&	-	&	-	&	-	&		&		\\
2FGLJ2330.6-3723 	&	0.27	&	2010/5/27-2010/6/4    	&	13.59 	&	-0.14 	&	-	&	-	&	-	&	45.70 	&	-	&	BL Lac-LSP 	\\
	&		&		&		&	-	&	-	&	-	&	-	&	-	&		&		\\
2FGLJ2347.0+5142 	&	0.044	&	2010/1/13-2010/1/22   	&	-	&	-	&	8.8	&	-	&	-	&	-	&	-	&	BL Lac 	\\
	&		&		&		&	-0.14 	&	11	&	-	&	-	&	44.00 	&		&		\\

\hline
\end{tabular}
\begin{quote}
References. 1: Chen et al.(2015); 2: Xiong et al.(2014b); 3: Ghisellini et al.(2014); 4: Chai et al.(2012); 5: Woo $\&$ Urry (2002); 6: Ghisellini et al.(2011); 7: Liang $\&$ Liu (2003); 8: Xie et al. (2007); 9: Celotti et al.(1997); 10: Shaw et al.(2012); 11: Zhang et al. (2012).
\end{quote}
\end{table*}

\end{document}